\documentclass[graybox, secnum]{svmult}

\usepackage{mathptmx}       
\usepackage{helvet}         
\usepackage{courier}        
\usepackage{type1cm}        
\usepackage{makeidx}         
\usepackage{graphicx}        
\usepackage{multicol}        
\usepackage[bottom]{footmisc}
\usepackage{hyperref}        
\usepackage{soul}            
\hypersetup{colorlinks=true,urlcolor=blue}
\usepackage[square,numbers]{natbib}
\makeindex             

\begin{document}
\title*{X-ray emission of massive stars and their winds}
\titlerunning{Massive stars and stellar winds}
\author{Gregor Rauw}
\institute{Space sciences, Technologies and Astrophysics Research (STAR) Institute, Universit\'e de Li\`ege, All\'ee du 6 Ao\^ut, 19c, B\^at B5c, 4000 Li\`ege, Belgium, \email{g.rauw@uliege.be}}
%
%
\maketitle
\abstract{Most types of massive stars display X-ray emission that is strongly affected by the properties of their stellar winds. Single non-magnetic OB stars have an X-ray luminosity that scales with their bolometric luminosity and their emission is thought to arise from a distribution of wind-embedded shocks. The lack of significant short-term stochastic variability indicates that the winds consist of a large number of independent fragments. Detailed investigations of temporal variability unveiled a connection between the photosphere and the wind: well-studied O-type stars exhibit a $\sim 10$\% modulation of their emission on timescales consistent with their rotation period, and a few early B-type pulsators display $\sim 10$\% modulations of their X-ray flux with the same period as their photospheric pulsations. Unlike OB stars, their evolved descendants (Wolf-Rayet stars and Luminous Blue Variables) lack a well-defined relation between their X-ray and bolometric luminosities, and several subcategories of objects remain undetected. These properties most likely stem from the combined effects of wind optical depth and wind velocity. Magnetic OB stars display an enhanced X-ray emission which is frequently modulated by the rotation of the star. These properties are well explained by the magnetically confined wind shock model and an oblique magnetic rotator configuration. Some massive binaries display phase-dependent excess emission arising from the collision between the winds of the binary components. Yet, the majority of the massive binaries do not show evidence for such an emission, probably as a consequence of radiative cooling of the shock-heated plasma. Finally, a growing subset of the Be stars, the so-called $\gamma$~Cas stars, feature an unusually hard and strong thermal X-ray emission that varies in a complicated manner over a wide range of timescales. Several scenarios  have been proposed to explain these properties, but the origin of the $\gamma$~Cas phenomenon remains currently one of the major unsolved puzzles in stellar X-ray astrophysics.}
\section*{Keywords} 
X-rays: stars; stars: early-type; stars: emission-line, Be; stars: massive; stars: winds, outflows; stars: Wolf-Rayet; binaries: close; line: profiles

\section{Introduction}
The upper left part of the Hertzsprung-Russell diagram is populated by massive ($M_* > 10$\,M$_{\odot}$), hot ($T_{\rm eff} \geq 20\,000$\,K) and luminous ($L_{\rm bol} \geq 5\,10^3$\,L$_{\odot}$) stars of spectral type O and B. These objects have a tremendous impact on their surroundings, notably via their huge UV luminosities and through the feedback of chemically enriched material and kinetic energy that they inject into the interstellar medium (ISM). This feedback occurs not only through the core-collapse supernova explosion that marks the end of their life, but happens already over their entire existence through a continuous mass-loss. Indeed, the intense UV radiation field accelerates the material in the atmosphere, leading to the formation of a dense and fast outflow: the stellar winds of main-sequence O-stars carry typical mass-loss rates in the range $\dot{M} = 2\,10^{-8}$ to $4\,10^{-6}$\,M$_{\odot}$\,yr$^{-1}$ and have asymptotic wind velocities of $v_{\infty} = 2000$ to $3000$\,km\,s$^{-1}$. Even higher mass-loss rates are observed in evolved massive stars, such as Wolf-Rayet (WR) stars or Luminous Blue Variables (LBVs). In parallel, a subgroup of rapidly rotating OB stars, the so-called Be and Oe stars, display (often) double-peaked Balmer emission lines in their optical spectra. These features are attributed to a viscous keplerian decretion disk located in the plane of the stellar equator. The circumstellar environment of massive stars (i.e.\ their stellar winds and the disks of Be stars) plays a key role in the interpretation of the X-ray emission of these objects. This environment is not only important because it absorbs part of the X-rays arising from very near the star, but it also plays an active role in the generation of the high-energy emission. 

Indirect evidence for the production of X-rays within the winds of massive stars came from {\it Copernicus} UV spectra which revealed lines from highly ionized species (O\,{\sc vi}, N\,{\sc v}, C\,{\sc iv}). In view of the effective temperatures of these stars, the existence of these ionization states could not be explained by photo-ionization effects, and was instead attributed to Auger ionization of the wind by the X-ray emission \citep{Cas79}. The confirmation that massive stars are indeed X-ray sources came when the {\it Einstein} satellite detected several OB and WR stars in the Carina\,OB1 and the Cygnus\,OB2 associations \citep{Sew79,Har79}. Since those early days, many satellites have contributed to a better understanding of the X-ray emission of massive stars. Spectacular progress has been achieved over the last two decades thanks to a fleet of X-ray satellites, notably the {\it Chandra} and {\it XMM-Newton} observatories (see the chapters by Wilkes \& Tananbaum and Schartel \& Santos-Lle\'{o} in Sect.\,III of this Handbook). Both satellites have collected CCD-resolution X-ray spectra for hundreds of massive stars, and, thanks to the High-Energy Transmission Grating (HETG) onboard {\it Chandra} and the Reflection Grating Spectrograph (RGS) onboard {\it XMM-Newton}, also high-resolution spectra for about two dozen of massive stars.    

This chapter deals with the X-ray properties of Population I massive stars of spectral types O, B, Be, WR and LBV. Single stars as well as non-degenerate massive binaries are considered, and the unusual X-ray emission of a subcategory of Be stars is highlighted. 

\section{X-ray emission from single massive stars}
\begin{figure}
  \begin{center}
    \resizebox{10cm}{!}{\includegraphics{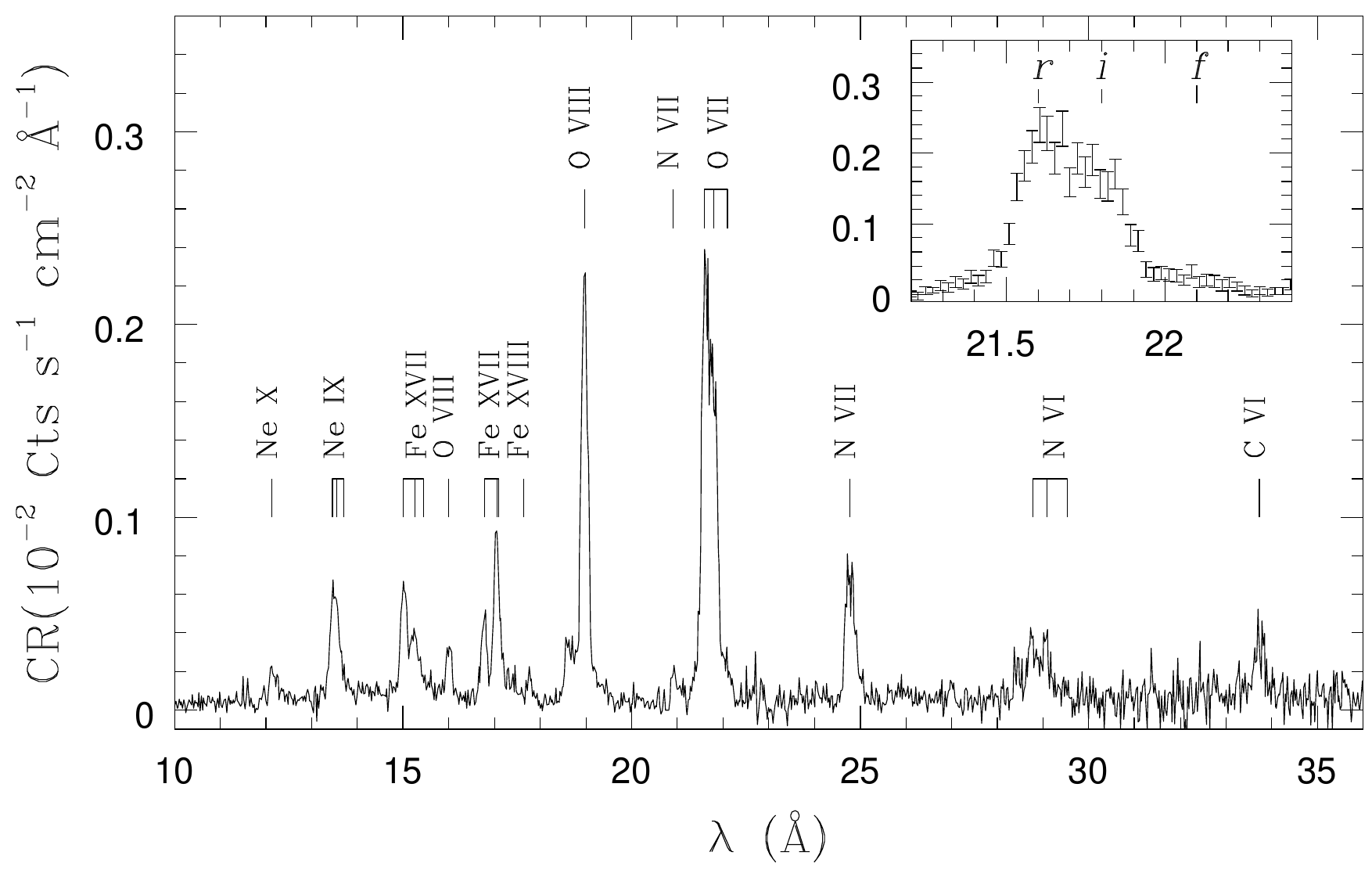}}
  \end{center}
  \caption{Combined {\it XMM-Newton} RGS\,1 and 2 spectrum of the O8\,III star $\lambda$~Ori. The most important emission lines are identified. The insert provides a zoom on the O\,{\sc vii} He-like triplet with the wavelengths of the resonace ($r$), intercombination ($i$) and forbidden ($f$) transitions indicated.\label{lamOri}}
\end{figure}

The X-ray spectra of massive stars mostly consist of emission lines of highly ionized species (see Fig.\,\ref{lamOri}). To first order, the overall X-ray spectra are rather well described by a multi-temperature optically-thin thermal plasma \citep[e.g.][]{Zhe07,Naz09,Coh21}. At first, it was suggested that the X-ray emission of single massive stars arises in a hot corona at the base of the wind \citep{Cas79}. Yet, photoelectric absorption by the overlying wind should then lead to a severe attenuation of the X-ray emission at energies below 1\,keV, which is not observed. This led to the elaboration of an alternative scenario where the X-rays arise from a distribution of shocks embedded inside the stellar winds \citep{Luc80}. Indeed, the line radiation pressure driving mechanism of the stellar winds is intrinsically unstable: a small seed perturbation of the velocity field grows as a result of the so-called line-deshadowing instability \citep[LDI,][]{Luc80,Owocki88,Feld97}. This leads to the development of shocks between parcels of wind material moving at different velocities. At these shocks, the kinetic energy of the flow is converted into heat and the shock-heated plasma subsequently cools through the emission of X-rays. Wind embedded shocks due to the LDI constitute nowadays the 'standard' scenario for explaining the intrinsic X-ray emission of single non-magnetic OB stars.  

\subsection{OB stars \label{OBstars}}
Soon after the discovery of X-ray emission from OB stars, it was noted that the observed X-ray luminosity of O-type stars (corrected for the absorption by the ISM) scales linearly with their bolometric luminosity \citep{Lon80,Pal81,Cas81} following $L_{\rm x}/L_{\rm bol} \simeq 10^{-7}$. This simple scaling relation has been subsequently confirmed and refined with {\it ROSAT}, {\it XMM-Newton} and {\it Chandra} data \citep[e.g.][]{Ber97,San06,Naz09,Naz11}. For B-type stars later than about B1, this relation breaks down and no clear correlation exists \citep{San06,Naz11,Rau15}. The majority of the O-star binaries have $L_{\rm x}/L_{\rm bol}$ values that agree well with those of presumably single O-stars \citep{Osk05,Naz09,Naz11,Rau15}. Only a few systems display significant X-ray overluminosities due to wind interactions which vary with orbital phase (see below).

At this point, it is important to stress that the $L_{\rm X}/L_{\rm bol}$ relation of O-type stars holds for the X-ray luminosities corrected for the sole ISM absorption. The X-ray spectra of O-type stars display the signature of additional photoelectric absorption due to the cool stellar wind material \citep{Naz09,Coh21}. Yet, the strong degeneracy between the temperature of the emitting plasma and the column density in the fitting procedure makes the actual wind column densities subject to large uncertainties. Therefore, estimates of the total X-ray emission generated in the wind, i.e.\ corrected for the wind absorption, are highly uncertain and show no well-defined trend with bolometric luminosity.

From the theoretical point of view, the $L_{\rm x}/L_{\rm bol}$ relation is a challenge. Indeed, the LDI scenario predicts that the X-ray luminosity of single massive stars should scale as $L_{\rm X} \propto \dot{M} \propto L_{\rm bol}^{1.7}$ for shocks in O-star winds where radiative cooling is efficient, whereas it should scale as $L_{\rm X} \propto \dot{M}^2 \propto L_{\rm bol}^{3.4}$ for shocks in the adiabatic regime expected in more tenuous B-star winds \citep{Owo13}. Both relations are steeper than the observational linear scaling relation. For shocks in the adiabatic regime, one possibility to recover the linear $L_{\rm x}/L_{\rm bol}$ relation is to assume that the hot plasma filling factor undergoes a radial decline according to $f \propto r^{-0.4}$ \citep{Owo99}. Yet, for radiative shocks, another explanation is required. In this case, the thin-shell instabilities that affect the radiative gas could induce mixing of hot and cool material, thereby leading to a softening and weakening of the observable X-ray emission. The canonical $L_{\rm x}/L_{\rm bol}$ relation can then be reproduced provided that the ensuing reduction of the X-ray emission scales with a power $m \simeq 0.4$ of the cooling length ratio \citep[defined as the ratio between the scaling distance of radiative cooling over the scale height of adiabatic cooling][]{Owo13}. One prediction of this scenario is that for the earliest and most luminous O-type stars, featuring dense winds, the $L_{\rm x}/L_{\rm bol}$ relation should saturate and decline with increasing $\dot{M}$ (hence with increasing $L_{\rm bol}$) because the optical depth of the winds to X-rays becomes significant. Evidence of such an effect has been reported for the O2\,If star HD93129A \citep{Coh11} and for the O4\,If$^+$ star HD15570 \citep{Rau16}.

Whilst it is likely that the shock-heated plasma in a stellar wind spans a roughly continuous range of temperatures, the X-ray spectra are usually fitted with models containing several discrete plasma components. For instance, analyzing CCD-type X-ray spectra of a large sample of O-type stars with models consisting of three plasma components yields typical temperatures of $kT = 0.2$, 0.6 and 2\,keV \citep{Naz09}. High-resolution X-ray spectra of O-stars basically agree with this picture, although the plasma temperatures can differ because of the differences in spectral response between gratings and CCD spectroscopy. For instance, the {\it XMM-Newton} RGS spectrum of the O4\,Infp star $\zeta$~Pup was modelled with four plasma components of $kT = 0.10$, 0.20, 0.40 and 0.69\,keV \citep{Her13}. Likewise, {\it Chandra}-HETG spectra of six OB stars were fitted with the sum of six thermal plasma components with fixed temperature evenly spaced in $\log{T}$ ($kT = $0.110, 0.187, 0.318, 0.540, 0.919, 1.56\,keV) to mimic a differential emission measure model \citep{Coh21}. 

\begin{figure}
  \begin{minipage}{7cm}
  \begin{center}
    \resizebox{7cm}{!}{\includegraphics{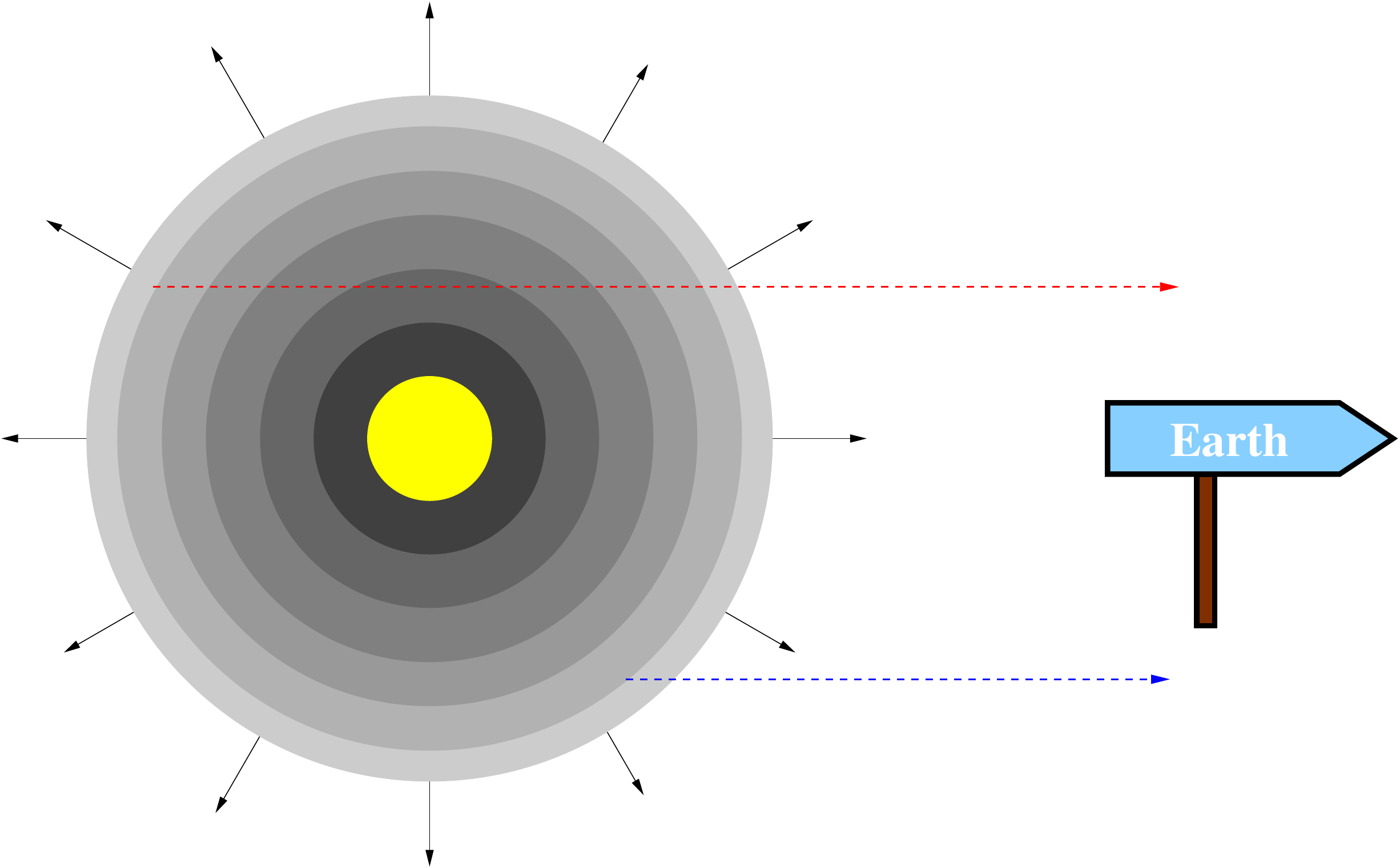}}
  \end{center}
  \end{minipage}
  \begin{minipage}{4.5cm}
  \begin{center}
    \resizebox{5cm}{!}{\includegraphics{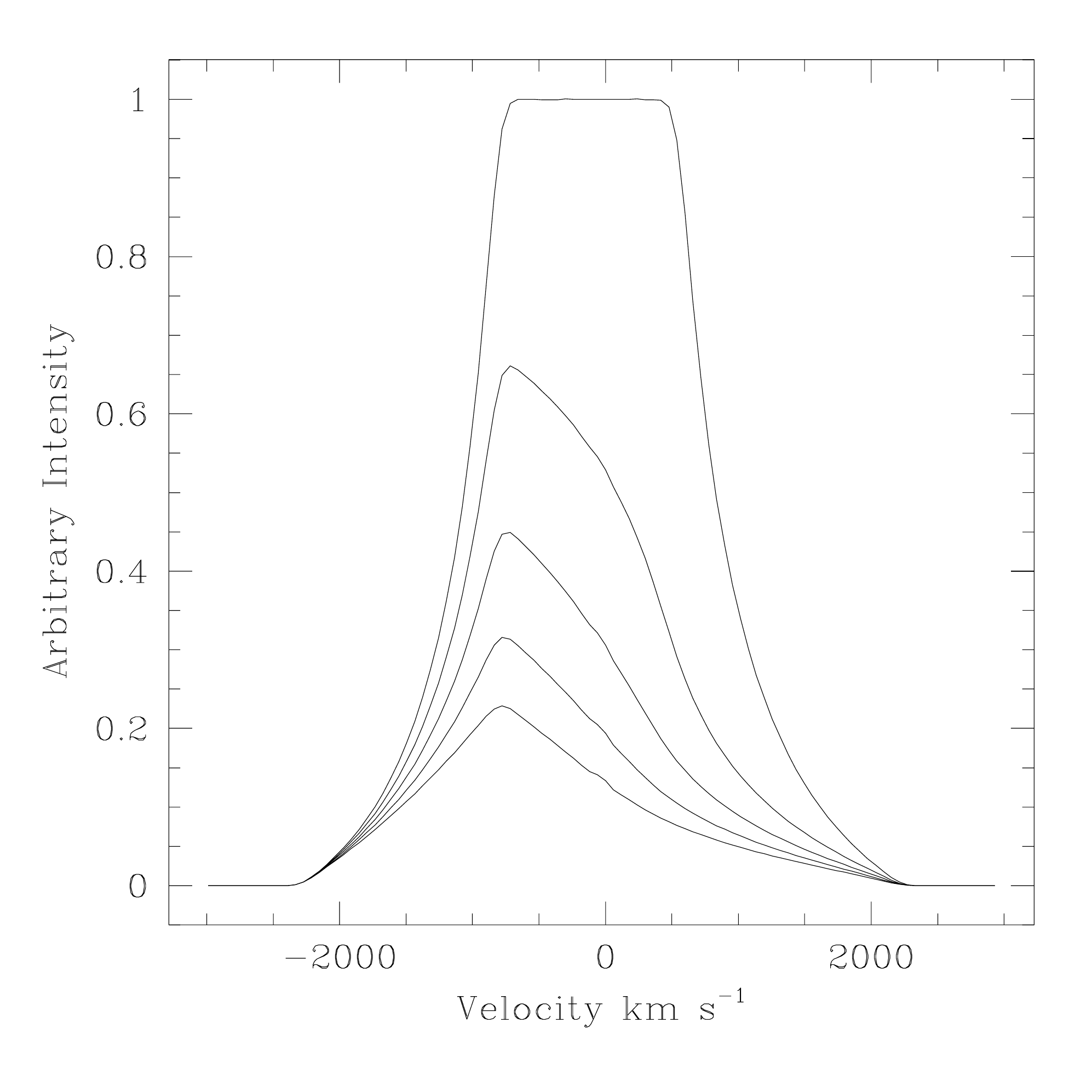}}
  \end{center}
  \end{minipage}
  \caption{Schematic illustration of the X-ray line formation process in the wind-embedded shock scenario. The line profiles on the right are computed assuming an emissivity that scales as $\rho^2$, $v_{\infty} = 2300$\,km\,s$^{-1}$, and an onset radius of the X-ray emission at 1.5\,$R_*$. Profiles with $\tau_{*, \lambda} = 0$ (flat-topped and symmetric profile), $0.5, 1.0, 1.5$ and $2.0$ (most skewed line profile) are shown.\label{skewed}}
\end{figure}
Observed X-ray line profiles broadly agree with the picture of a distributed X-ray emission attenuated by absorption by the cooler wind. This model was first elaborated for extreme UV lines \citep{Mac91} and then adapted to X-ray lines \citep{Owo01}. Assuming a spherically symmetric wind seen by an external observer, we can take advantage of the azimuthal symmetry and describe the problem in the $(p,z)$ cylindrical coordinates, where $p$ is the radial coordinate and the observer is located at $z \rightarrow \infty$. The optical depth from a given position inside the wind to the observer is then given by
\begin{equation}
  \tau_{\lambda}(p,z)  = \int_z^{\infty}\kappa_{\lambda}\,\rho(p,z')\,dz' =  \tau_{*, \lambda}\,\int_z^{\infty}\frac{R_*\,dz'}{r'^2\,(1-R_*/r')^{\beta}}
    \label{Eq2}
\end{equation}
where $\kappa_{\lambda}$ is the opacity at wavelength $\lambda$, $\rho$ is the wind density, $R_*$ is the stellar radius, and $r' = \sqrt{p^2 + z'^2}$. The wind is assumed to follow a $v(r') = v_{\infty}\,\left(1 - \frac{R_*}{r'}\right)^\beta$ velocity law, and
\begin{equation}
  \tau_{*, \lambda} = \frac{\kappa_{\lambda}\,\dot{M}}{4\,\pi\,v_{\infty}\,R_*}
\end{equation}
with $\dot{M}$ the stellar mass-loss rate. Here, Eq.\,\ref{Eq2} implicitly assumes $\kappa_{\lambda}$ to be constant as a function of position $r'$ in the wind. 

As illustrated in Fig.\,\ref{skewed}, wind absorption is expected to produce shifted and skewed line profiles. The optical paths of two representative photons, emitted from the same shell of shock-heated material but from opposite sides, are shown by the dashed lines. The photon coming from the rear side of the wind is emitted by material moving away from the observer and appears therefore red-shifted in the observer's frame of reference. This photon travels a long way through rather dense material before getting out of the wind, and suffers thus attenuation by a rather large optical depth. On the other hand, the photon coming from the front side is emitted by material moving towards the observer and appears thus blue-shifted to the observer. The column of material that this photon has to cross before leaving the wind is quite modest, implying a small optical depth. The morphology of the resulting line profiles depends on the value of the wind opacity via $\tau_{*, \lambda}$. For small values of $\tau_{*, \lambda}$, the lines are broad and display a box-like shape (see right panel of Fig.\,\ref{skewed}). For increasing values of $\tau_{*, \lambda}$, the lines become more and more skewed since the red-shifted photons suffer from heavier line-of-sight absorption than the blue-shifted photons.

Because of the wavelength dependence of $\kappa_\lambda$, one expects quite large variations of $\tau_{*, \lambda}$ over the X-ray band which should result in differences in shape between the spectral lines at longer and shorter wavelengths. This dependence could in principle be used to infer the stellar mass loss rate \citep{Leu13,Coh14a,Coh20}. Yet, the observed X-ray lines of O-type stars generally show rather little asymmetry \citep[e.g.][]{Coh14a}. This could be a manifestation of clumping of the cool wind which implies that the simple absorption treatment described above must be modified. For instance, a fragmented cool wind leads to porosity effects when the line of sight passes in between the clumps, allowing the photons to escape more easily. The resulting line profiles depend on the optical thickness and on the geometry of individual clumps \citep{Feld03,Osk04,Osk06,Owo06,Sun12,Her12}. This leads to the definition of an effective absorption coefficient \citep{Ignace}:
\begin{equation}
  \kappa_{\rm eff}\,\rho = n_{\rm cl}\,A_{\rm cl}\,(1 - \exp{(-\tau_{\rm cl})}) = \frac{1 - \exp{(-\tau_{\rm cl}})}{\tau_{\rm cl}}\,\kappa\,\overline{\rho}
\end{equation}
where $n_{\rm cl}$, $A_{\rm cl}$ and $\tau_{\rm cl}$ are the number density, cross-section and optical depth of the clumps. The latter quantity is expressed as $\tau_{\rm cl} = \kappa\,\overline{\rho}\,h$ where $\overline{\rho}$ is the average density and $h = \frac{1}{n_{\rm cl}\,A_{\rm cl}}$ is the porosity length \citep{OGS04}. For optically thin clumps the $\frac{1 - \exp{(-\tau_{\rm cl}})}{\tau_{\rm cl}}$ corrective factor reduces to 1, whilst it becomes $1/\tau_{\rm cl}$ for optically thick clumps. In the latter case, the porosity of the wind significantly affects the line shape: large porosity lengths make the wind more transparent to X-rays, implying more symmetric line profiles. As far as the clump geometry is concerned, isotropic (i.e.\ spherical) and anisotropic (i.e.\ flat, pan-cake shaped) fragments have been considered \citep{Feld03,Osk04,Her12}. Whilst hydrodynamical simulations predict relatively flat fragments, it was found that anisotropic clumps yield rather poor fits to the observed line profiles of $\zeta$~Pup whilst a moderate level of isotropic porosity could not be ruled out \citep{Her13,Leu13}. In view of the degeneracy between the value of the mass-loss rate, the porosity parameter and the geometry of the fragments that make up the cool wind, the question as to whether or not high-resolution X-ray spectra can provide self-consistent and independent determinations of $\dot{M}$ remains debated \citep{Her12,Leu13,Osk16}. Nonetheless, high-resolution X-ray spectra offer a precious ingredient of multi-wavelength analyses allowing to perform consistency checks of mass-loss rates and properties of the fragments inferred from optical and UV spectra \citep[e.g.][]{Her13,Osk16}.

High-resolution X-ray spectra can be used to estimate the radius at which the X-ray emission starts inside the wind. Indeed, the helium-like ions (Mg\,{\sc xi}, Ne\,{\sc ix}, O\,{\sc vii}, N\,{\sc vi}) display triplets in the RGS and HETG energy domain that consist of a resonance ($r$), an intercombination ($i$) and a forbidden ($f$) line. In the presence of either a high plasma density or a strong UV radiation field, the upper level of the $f$ transition is depopulated at the benefit of the upper level of the $i$ line. In stellar winds of massive stars, the radiation field dominates over density effects and the ratio $f/i$ offers a sensitive diagnostic of the dilution of the photospheric UV radiation at the location of the X-ray plasma \citep{Por01}. In the spectra of O-type stars, the $f$ line is strongly suppressed whilst the $i$ component is strong (see e.g.\ Fig.\,\ref{lamOri}). This indicates that the X-ray emission arises in the inner parts of the wind, typically at $\sim 0.5\,R_*$ above the photosphere, where the photospheric UV radiation is strong \citep{Leu06,Leu07,Osk06}. Whilst most studies that explore this diagnostic assume that each $f i r$ triplet arises from a single-temperature plasma, reality is more complicated as several plasma components with different onset radii can contribute to the formation of these triplets \citep{Her13}. However, the conclusion that the X-ray emission starts already relatively close to the photosphere remains valid.  

So far, numerical LDI hydrodynamic simulations that include the energy equation, needed to estimate the X-ray emission, are restricted to 1-D \citep{Feld97}. These simulations predict large-amplitude stochastic X-ray variability either as a result of variations of the emission measure of the hot gas or from fluctuations of the absorbing column density due to clumps of cool material along the line of sight. These large amplitude variations are clearly an artefact due to the 1-D nature of the calculations. Indeed, the large sample of {\it XMM-Newton} spectra of $\zeta$~Pup indicates that any stochastic short-term variability has an amplitude $< 1$\%, i.e.\ comparable to, or even smaller than, the Poisson noise of the data. This result translates into a lower limit on the number of independent X-ray emitting and X-ray absorbing fragments in the stellar wind of at least $10^5$ at any given time \citep{Naz13}.

Beside the small-scale fragments, stellar winds can also host large-scale structures due to so-called co-rotating interaction regions \citep{Cra96,Lob08}. As a result of the stellar rotation, the trajectories of radially ejected wind material in an external frame of reference are spirals with a curvature set by the ratio between the stellar rotational velocity and the wind velocity. When the star features a bright spot at its surface (e.g.\ as a consequence of a localized magnetic field generated in a thin subsurface convective layer \citep{Can09}), this alters the wind velocity locally, resulting in a spiral-like interaction region due to the collision of material moving at different wind velocities. The rotation of the star then leads to a corotating interaction region (CIR) that modulates the column density of the material along the line of sight. Evidence of recurrent modulations of the X-ray fluxes at the $5$ -- $10$\% level on timescales of days, likely due to CIRs, has accumulated for several objects. Following the discovery of variability on timescales of days in the case of $\zeta$~Pup \citep{Naz13,Naz18}, an extensive 813\,ks {\it Chandra} campaign on this star spread over about one year led to the detection of two periodicities: a highly significant modulation at 1.78\,days and a more marginal signal near 5 -- 6 days \citep{Nic21}. The 1.78\,day period is also present in space-borne optical photometry of $\zeta$~Pup and was interpreted as the star's rotational period \citep{Nic21}. Other stars displaying modulations of the X-ray emission on similar timescales have been found. For the O7\,III(n)((f)) giant $\xi$~Per, coordinated {\it XMM-Newton} X-ray and {\it HST}-STIS UV spectroscopy revealed evidence for a modulation on a 2.086\,day timescale \citep{Mas19}. For the O6\,Infp star $\lambda$~Cep, coordinated {\it XMM-Newton} and optical spectroscopy suggest the existence of a 4.1\,day modulation both in the H$\alpha$ equivalent width (EW) and in the X-ray flux \citep{Rau15b}.

An open question is how exactly a CIR impacts the level of the observed X-ray emission. This could happen either as a result of extra X-ray emission produced by the shocks of the CIR itself or through the density enhancement associated with the CIR leading to a modulation of the column-density. It remains to be seen whether or not the velocity jumps in the CIRs are sufficient to provide a significant X-ray emission. As an alternative, additional X-ray emission could be directly associated with the localized magnetic field responsible for the spot that generates the CIR.
\begin{figure}
  \begin{center}
    \resizebox{8cm}{!}{\includegraphics{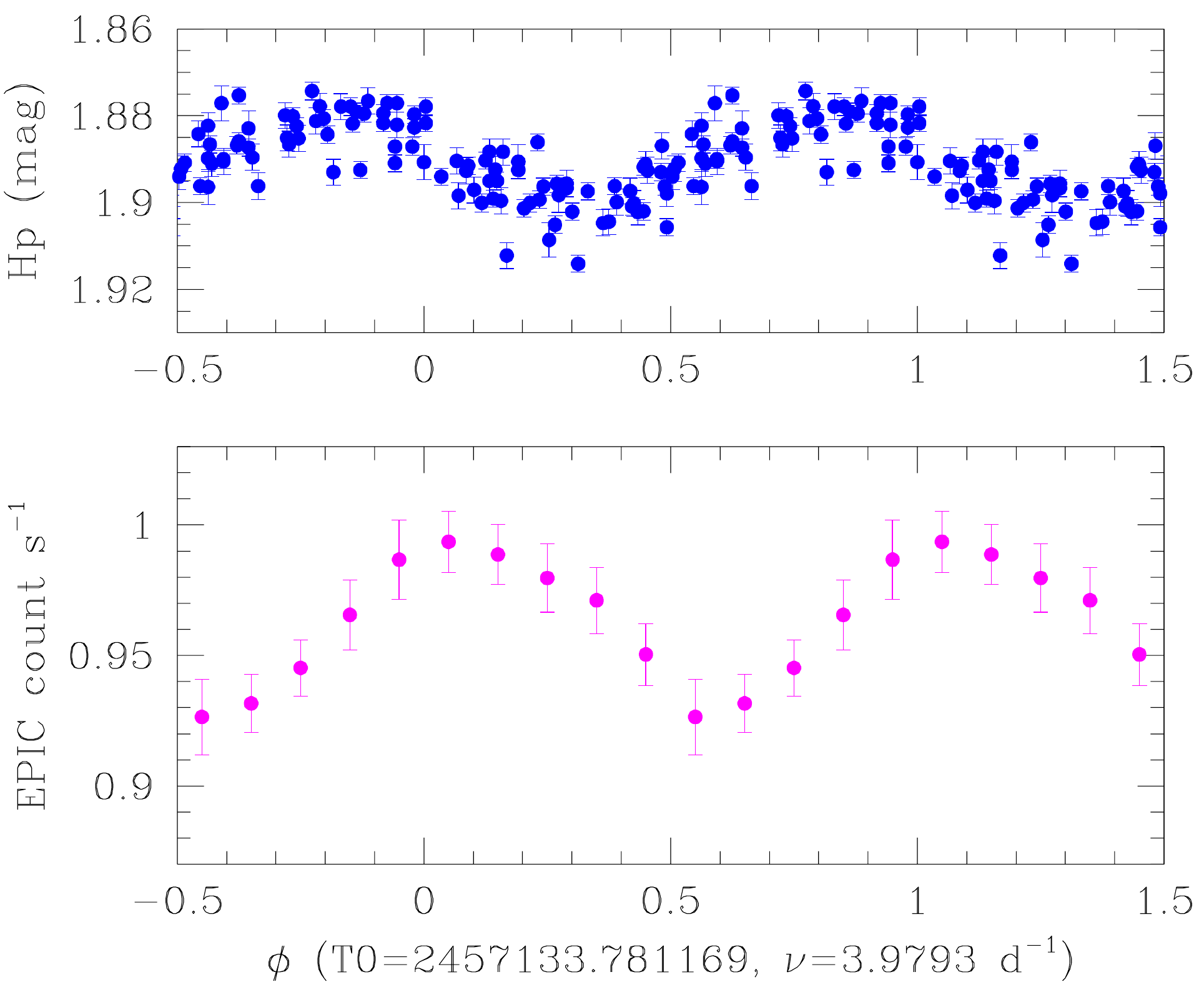}}
  \end{center}
  \caption{Optical and X-ray pulsations of $\beta$~CMa (B1\,II-III). The top panel illustrates the {\it Hipparcos} photometry, retaining only points with uncertainties of less than 0.005\,mag, folded with the 6.03\,h pulsation period \citep{Sho06}. The bottom panel illustrates the EPIC count rate from {\it XMM-Newton} revolution 2814 folded with the same period and zero point \citep{Caz17}.\label{betCMa}}
\end{figure}

Further evidence for a direct connection between the photospheric conditions and the X-ray emission comes from the detection of X-ray pulsations in (at least) two stars of the class of $\beta$~Cep pulsators: $\xi^1$~CMa \citep{Osk14} and $\beta$~CMa \citep{Caz17}. These X-ray pulsations (at the $\sim 10$\% level, see Fig.\,\ref{betCMa}) were found to occur on the same period as the pulsations seen in the optical light curves. It has to be stressed though that at the current level of sensitivity, this does not seem to be a general feature of $\beta$~Cep stars. Indeed, searches for such X-ray pulsations in several other $\beta$~Cep stars failed to reveal a significant modulation \citep{Osk15}. How exactly the connection between the photosphere and the hot plasma operates remains to be established.

\subsection{Evolved massive stars}
Though the details of the evolution of massive stars are still not fully understood, it is commonly accepted that most Wolf-Rayet stars are the evolved descendants of O-type stars \citep[for a review see][]{Cro07}. WR stars are classified into three sequences: WN stars showing strong He and N emission lines\footnote{Hydrogen-rich WN stars are very massive ($M_* \geq 65$\,M$_{\odot}$) O-stars with strong winds rather than classical hydrogen-poor WN stars \citep{Cro07}.}, WC stars displaying prominent He and C emissions, and WO stars with spectra dominated by emissions of He, C and O. The spectra of these stars are interpreted as the result of a strong mass-loss process that removed the outer H-rich layers of the star, thereby unveiling the products of the CNO nucleosynthesis cycle (for WN stars) or the products of He-burning (for WC and WO stars). WR stars have significantly larger wind mass-loss rates than O-type stars, typically in the range $5\,10^{-6}$ to $2\,10^{-5}$\,M$_{\odot}$\,yr$^{-1}$. Their wind velocities range between $\sim 700$\,km\,s$^{-1}$ for WN9 stars and up to $6000$\,km\,s$^{-1}$ for WO stars \citep{Cro07}.   

How an O-star loses the mass to become a WR star is one of the key unknowns of massive star evolution, but for the most massive stars, this probably happens through a short-lived LBV phase \citep{Hum94}. Such LBVs have $\dot{M}$ up to $10^{-4}$\,M$_{\odot}$\,yr$^{-1}$ and $L_{\rm bol}$ up to $10^6$\,L$_{\odot}$. They often undergo photometric and spectroscopic variability with the most extreme mass ejection events occurring during giant eruptions. Their winds are generally rather slow.

Early X-ray observations of WR stars with {\it Einstein} suggested that the X-ray brightest WR stars are often WR + O binaries, whilst single WR stars are usually faint X-ray sources, and WN stars are on average brighter X-ray emitters than WC stars \citep{Pol87}. Unlike OB stars, presumably single WR stars display a huge scatter in their $L_{\rm x}/L_{\rm bol}$ value \citep{Wes96,Ign99,Osk05}, and the level of observable X-rays depends on the spectral type. With the current generation of X-ray facilities, presumably single early WN (WNE) stars are detected as moderate X-ray emitters \citep{Osk16b}, whilst  most late WN (WNL) and WC stars are X-ray faint or even dark \citep{Osk03,Gos05,Ski12,Osk16b}. Quite remarkably, the WO2 star WR~142 was detected with $\log{L_{\rm X}/L_{\rm bol}} \simeq -8.1$ \citep{Osk09}. 
At first sight, the non-detection of some categories of WR stars could be related to the optical depth of their dense winds which leads to values of the radius $R_1$ where the radial optical depth reaches unity of several thousand stellar radii. This explanation works rather well for the X-ray darkness of single WC stars which have very opaque winds \citep{Osk03}. The first presumably single WC star detected in X-rays is the WC4 star WR~144 which has $\log{L_{\rm X}/L_{\rm bol}} \simeq -8.8$ \citep{Rau15}. Yet, the wind optical depths alone cannot account for all non-detections. Indeed, the absence of X-rays of the WN8h star WR~40 \citep{Gos05} cannot be explained that way, since its wind should be more X-ray transparent than those of most WNE stars which are however detected.

A directly related question concerns the nature of the X-ray emission process in WR stars. A natural candidate is the LDI, which is thought to be responsible for the clumpiness of WR winds, though multiple photon scattering in the dense winds of WR stars is expected to lower the impact of LDI \citep{Gay95}. A promising attempt to unify the X-ray heating mechanism of OB and WR stars, simultaneously explaining the non-detections, relies on the interplay of shocks, wind optical depth, wind velocity and plasma cooling efficiency \citep{Gayley}. In this scenario, plasma is heated in the wind acceleration zone through shocks between the fast wind and slower clumps. The shock-heated plasma is then advected outwards. If it reaches the radius of optical depth unity on a timescale shorter than the plasma cooling time, then X-rays are observable. The X-ray emergence efficiency is thus controlled by the balance between the generation of hot X-ray emitting plasma and the wind's capability to advect it rapidly to large radii on the one hand, and the combination of X-ray absorption and plasma cooling on the other hand \citep{Gayley}. For WNL stars, such as WR~40, which have winds with low $v_{\infty}$ velocities, X-rays emitted by a putative hot gas would thus remain hidden because the gas cannot be advected outside the radius $R_1$ before its temperature drops below the threshold for X-ray emission. The WNE stars, as well as the WC4 star WR~144 and the WO2 star WR~142 have much faster winds, thus explaining their detection.  

The only presumably single WR star bright enough to be studied at high spectral resolution with current X-ray telescopes is the WN4 star WR~6 \citep{Osk12,Hue15}. The RGS and HETG spectra of WR~6 showed that the X-rays arise from very far out in the wind (out to 1000\,R$_*$), i.e.\ from near $R_1$, as determined from $f i r$ line ratios of He-like ions. The line profiles are consistent with the expected morphology of X-ray lines originating from a uniformly expanding spherical wind of high X-ray-continuum optical depth \citep{Hue15,Ignace}.
However, when interpreting the X-ray spectrum of WR~6, one must keep in mind the uncertain nature of this object. Indeed, WR~6 is notorious for its well-established photometric period of 3.765 days seen in optical and UV data. Though the period is stable, the modulation is highly variable in amplitude and shape. The origin of this phenomenon (CIR in a single, rotating star or binarity) remains debated \citep{StL18,Koe20}. Yet, whilst the X-ray flux of WR~6 clearly varies, these variations do not exhibit the 3.765\,day period \citep{Osk12,Hue15}. 

A survey of X-ray emission of Galactic LBVs with {\it XMM-Newton} and {\it Chandra} clearly demonstrated that these objects are not intrinsically bright X-ray emitters \citep{Naz12}: out of 31 LBVs and LBV candidates for which X-ray data existed, only four were detected and two more had doubtful detections. The non-detection of P~Cyg yields a stringent $\log{L_{\rm x}/L_{\rm bol}} \leq -9.4$ \citep{Naz12}. This star has a dense ($\dot{M} = 2\,10^{-5}$\,M$_{\odot}$\,yr$^{-1}$) and slow ($v_{\infty} = 185$\,km\,s$^{-1}$) wind. In the generalized LDI scenario, the resulting shocks are expected to be weak. Moreover, the wind has a huge optical depth and the wind velocity is not sufficient to advect the putative hot material beyond the $R_1$ radius before it cools down. Conversely, the brightest detections concern $\eta$~Car and Schulte~12. The former is a 5.5\,yr highly eccentric colliding wind binary, indicating that its X-ray emission most probably arises from the wind interaction rather than inside the wind of the LBV \citep{Oka08,Par11}. As to Schulte~12, this star displays a bright ($\log{L_{\rm x}/L_{\rm bol}} = -6.1$) and hard ($kT \sim 2$\,keV) emission \citep{Rau11,Caz14} that was shown to undergo a 108\,day modulation which is reminiscent of a colliding wind binary \citep{Naz19}. Yet, the optical spectra and photometry of this star do not vary as regularly as the X-ray data, but rather undergo variations on timescales of 50 -- 100\,days, which are more likely to result from pulsations. Moreover, the {\it Gaia} parallax of this star suggests that it is a normal B5\,Ia supergiant \citep[i.e.\ not a genuine LBV candidate,][]{Naz19}. Whilst the origin of the X-ray emission of Schulte~12 remains mysterious, it actually seems unrelated to the LBV phenomenon. 

An important issue concerns the evaluation of $L_{\rm X}$ of evolved massive stars. X-ray spectra of massive stars are frequently modelled by a combination of optically thin thermal plasma models absorbed by a column of cool material consisting of contributions from a local (i.e.\ wind) and an ISM component. To evaluate the intrinsic, unabsorbed, X-ray luminosity some authors correct the X-ray fluxes for the entire column density \citep[e.g.][]{Ski12,Zhe12}, whilst others correct only for the sole ISM absorption \citep[e.g][]{Rau15,NGM21}. The former approach might at first sight look more relevant since it aims at evaluating the actual power that is emitted as X-rays, but there are several important caveats. First of all, the models used to adjust the X-ray spectrum are subject to a high level of degeneracy: quite often the same spectrum can be adjusted equally well with either a higher temperature plasma absorbed by a lower column density or a lower temperature model seen through a higher column. This degeneracy leads to huge uncertainties, up to several dex, on the actual corrections to be applied to infer the intrinsic fluxes. The uncertainties are amplified by the fact that the wind column densities of evolved massive stars can be very large \citep[e.g][]{ZheSki15}, thereby leading to huge errors on the ISM + wind absorption-corrected fluxes. Moreover, the meaning of such fluxes corrected for the total column strongly depends on the actual origin of the X-ray emission, and thus the geometry of the emitting region, which is often unknown. Indeed, considering that the X-rays arise from a distribution of shocks in the winds of a single star, each cell of X-ray plasma is seen through a different value of the column density, and the usage of a single value of the wind column density can at most provide a mean value. Moreover, in such a situation, the wind itself is the source of the X-ray emission, and, because the energy of X-rays that are absorbed by the wind is simply re-injected into the wind, the only physically meaningful X-ray luminosity is the value that escapes from the wind into the interstellar medium. Special care is required for non-detections: upper limits on the X-ray flux should be evaluated only by correcting for the ISM absorption and cannot be compared with the 'fully corrected' fluxes resulting from spectral fits \citep{Ski12}.


\subsection{Magnetic massive stars \label{magfields}}
Spectropolarimetric surveys of massive stars in our Galaxy showed that $\sim 7$\% of the OB-stars display a strong (kG), large-scale, mostly dipolar, magnetic field \citep{Gru17}. The interplay between this field and an otherwise spherically-symmetric wind mass-loss leads to the formation of magnetically confined wind shocks \citep[MCWS,][]{Bab97,udD02,udDYN}. Within the Alfv\'en radius ($R_A$), i.e.\ within the distance from the star where the dynamics of the gas is ruled by the magnetic field, the ouflow is channeled along the magnetic field lines. The efficiency of the magnetic field to confine the wind is expressed via the wind magnetic confinement parameter \citep[][see also chapter by ud-Doula \& Owocki]{udD02}:
\begin{equation}
  \eta_* = \frac{B_{\rm eq}^2\,R_*^2}{\dot{M}_{B=0}\,v_{\infty}}
\end{equation}
where $B_{\rm eq}$ is the field strength at the magnetic equator, and the mass-loss rate $\dot{M}_{B=0}$ and wind velocity in the denominator refer to a fiducial situation of the same star but without a magnetic field. For magnetic O-type stars, $\eta_*$ has values of $\simeq $10 -- 100, whereas this parameter typically reaches $10^4$ -- $10^6$ for magnetic B-type stars \citep{Pet13,udD14,udDYN}.

As a result of this confinement, the wind outflows arising from opposite footpoints of closed magnetic field loops collide near the magnetic equator. The MCWS convert the kinetic energy of the inflowing cool gas into heat, leading to the emission of X-rays \citep{Bab97,udDYN}. 
\begin{figure}
  \begin{minipage}{7cm}
  \begin{center}
    \resizebox{7cm}{!}{\includegraphics{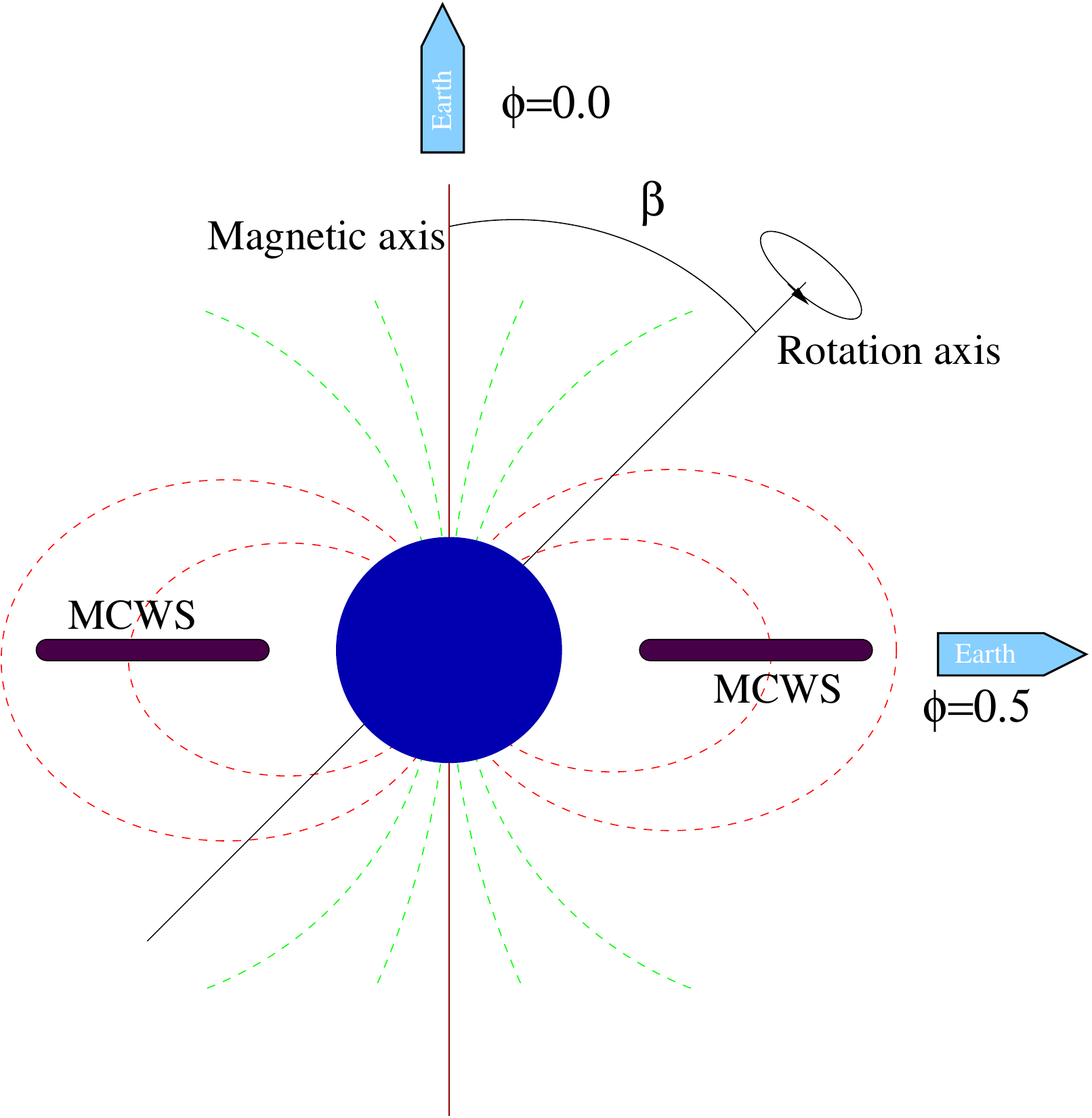}}
  \end{center}
  \end{minipage}
  \begin{minipage}{4.5cm}
  \begin{center}
    \resizebox{4.5cm}{!}{\includegraphics{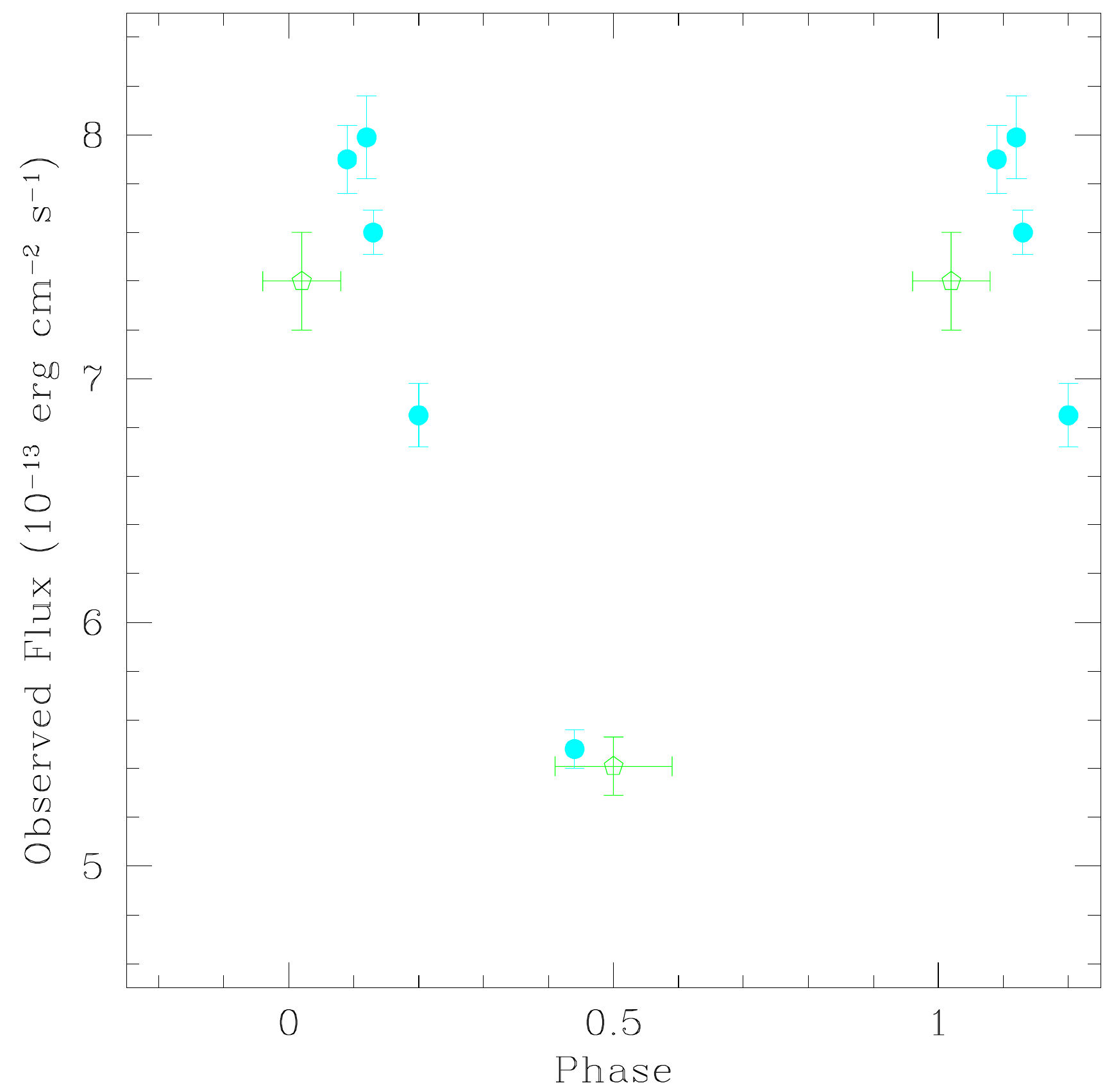}}
    \resizebox{4.5cm}{!}{\includegraphics{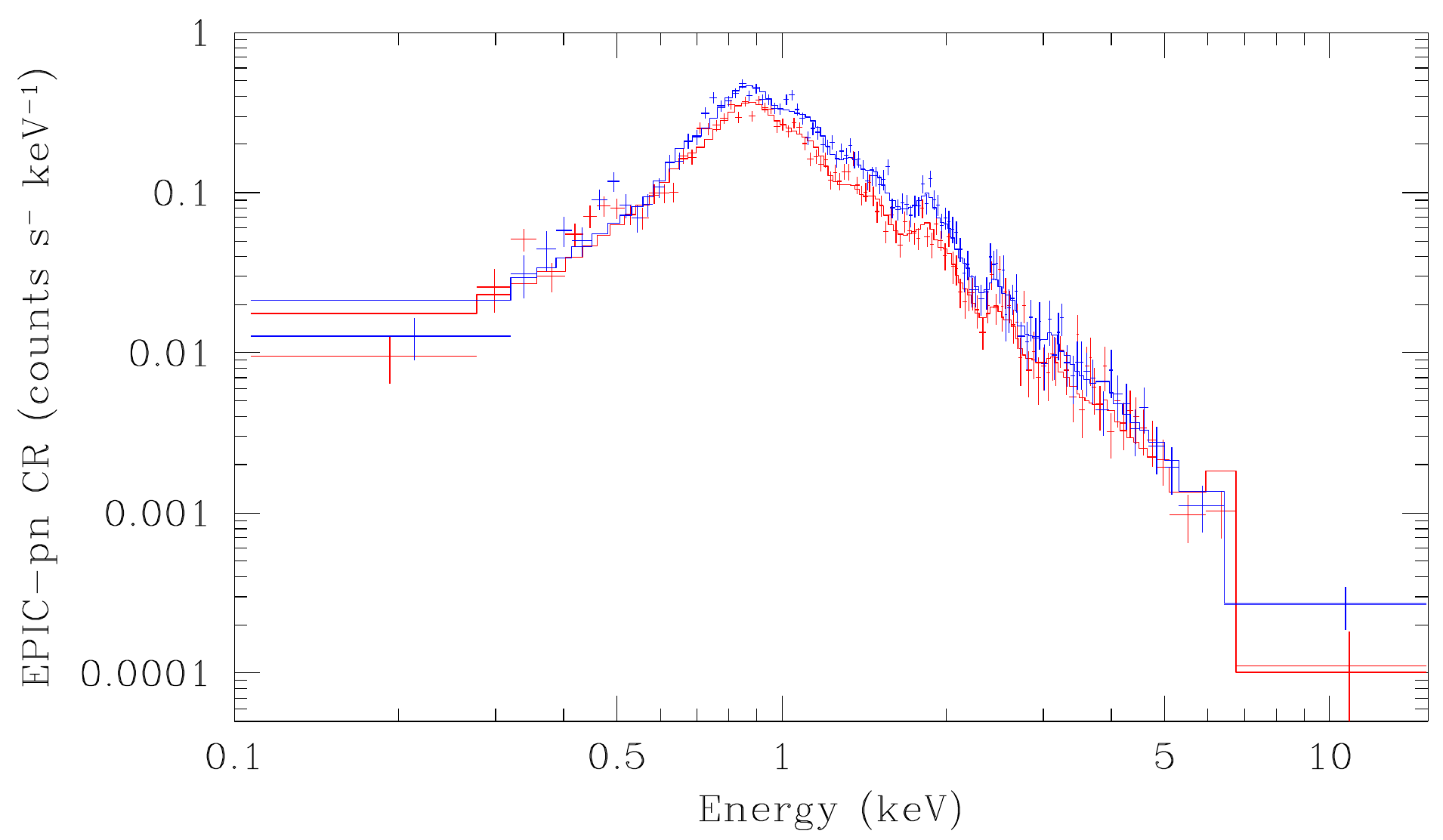}}    
  \end{center}
  \end{minipage}
  \caption{Left panel: schematic illustration of the oblique magnetic rotator model. Top right panel: variation of the X-ray flux (between 0.5 and 10\,keV) over the 537.6\,day rotation cycle of HD191612 \citep{Naz16}. Phase 0.0 corresponds to the maximum strength of the H$\alpha$ emission. Cyan dots and green pentagons correspond respectively to {\it XMM-Newton} and {\it Chandra} data. For this star, the obliquity angle $\beta$ and the inclination $i$ of the rotation axis with respect to our line of sight are such that $\beta + i \simeq 95^{\circ} \pm 10^{\circ}$ with $\beta \simeq i$ \citep{Sun12b}. Bottom right panel: comparison between two {\it XMM-Newton} EPIC-pn spectra of HD191612 obtained at phases 0.44 (red symbols) and 0.13 (blue symbols).\label{HD191612}}
\end{figure}
The nature of the interaction between the wind and the magnetic field further depends on the stellar rotation rate (for details see the chapter by ud-Doula \& Owocki). In a rotating magnetic star, the torques from the magnetic field maintain the wind material in rigid-body co-rotation out to $R_A$. The Kepler co-rotation radius ($R_K$) is defined as the radius where the centrifugal force associated with the rigid rotation balances the inward pull from gravity. If $R_A < R_K$, the star has a dynamical magnetosphere (DM), whereas stars with $R_A > R_K$ have a centrifugal magnetospheres \citep[CM,][]{Pet13}. In CMs, the centrifugal force leads to the formation of a dense, stable, rigidly rotating circumstellar disk. The trapped material accumulates, and is evacuated via episodic centrifugal ejection events. During such an ejection, magnetic reconnection events are expected \citep{Tow05,udD06}. They could correspond to the X-ray flares that were observed in the B2\,IV-Vp star $\sigma$~Ori\,E \citep[e.g.][]{Gro04,Ski08}, though the association of these flares with the massive star is still debated \citep{udDYN}.

When the magnetic field axis is inclined with respect to the rotation axis, the viewing angle onto the magnetosphere changes as a function of rotation phase (see Fig.\,\ref{HD191612}), thereby leading to periodic variations of the optical and UV emission line profiles and fluxes, as well as of the X-ray flux. This oblique magnetic rotator model was used to explain the periodic multi-wavelength variability of the O7\,V star $\theta^1$~Ori\,C \citep[e.g.][]{Sta96,Bab97,Ste05,Gag05}, well before the direct detection of its 1.1\,kG magnetic field. Rotational modulation of the X-ray emission and spectral hardness have subsequently been observed in a number of magnetic O-type stars. For instance, in the case of the O7f?p star NGC1624-2, which has the strongest magnetic field ($\sim 20$\,kG) observed in any O-type star so far, the X-ray flux varies by 30\% between the times of maximum and minimum H$\alpha$ emission \citep{Pet15}. Variations of a similar amplitude are observed for the O6-8f?p star HD191612 \citep[][see also Fig.\,\ref{HD191612}]{Naz16}. The variations of HD191612's X-ray spectrum mostly consist in a global scaling of the emission, probably as a result of occultation effects. By contrast, no rotational modulation of the X-ray emission was found for the B0.2\,V star $\tau$~Sco, despite a rather complex magnetic topology \citep{Ign10}. In the case of $\sigma$~Ori\,E, the absence of rotational modulation is explained by the huge size of its magnetosphere which minimizes the impact of  any occultation effects on the observed level of the emission \citep{YN14}. 

{\it XMM-Newton} and {\it Chandra} observations of 40 magnetic OB stars, spanning a wide range in $\eta_*$ values, showed that the X-ray luminosity strongly correlates with $\dot{M}$ \citep{YN14}. For most stars, the relation between $L_{\rm X}$ and $\dot{M}$ closely follows semi-analytical models of DMs \citep{udD14} scaled-down by a factor 0.1. The largest deviations are found for a few stars with CMs and for the DM of NGC1624-2. The huge magnetosphere of the latter star absorbs the majority of the X-rays, thus leading to a lower level of emission than expected. Whilst the emission of magnetic O-stars is on average somewhat harder than that of non-magnetic O-stars, there is a large scatter and there is no clear relationship between the hardness and the X-ray overluminosity \citep{YN14}. Finally, the X-ray spectral lines of OB stars with MCWS are expected to be narrow because they form in a slowly moving (confined) plasma. Observations indeed reveal rather narrow lines with widths significantly lower than those observed for single stars with wind-embedded shocks \citep[][and references therein]{udDYN}. 

\section{Massive binaries \label{CWB}}
When the X-ray emission from massive stars was discovered with {\it Einstein}, some binary systems were among the brightest sources \citep{Pol87,Chl91}. Their apparent overluminosity was attributed to an excess X-ray emission arising in colliding wind binaries (CWBs) \citep{Che76,Pri76}. Indeed, the head-on collision of highly supersonic winds leads to the formation of an interaction region between the two stars. This interaction region is contained between two oppositely-faced strong hydrodynamic shocks and the post-shock regions of the two winds are separated by a contact discontinuity \citep{SBP}. The shape and location of the contact discontinuity between the two winds are set by the wind momentum ratio parameter \citep{Can96}
\begin{equation}
  \eta = \frac{\dot{M_1}\,v_1}{\dot{M_2}\,v_2}
\end{equation}
where the 1 and 2 subscripts stand for the primary and secondary star, and $v_1$ and $v_2$ are the pre-shock wind velocities. At the shock front, the kinetic energy normal to the shock is converted into heat and the temperature of the post-shock plasma immediately behind the shock is given by 
\begin{equation}
  k\,T = \frac{3}{16}\,m_{\rm part}\,v_{\perp}^2
  \label{kinetic}
\end{equation}
where $m_{\rm part}$ and $v_{\perp}$ are the mass of a wind particle and the pre-shock wind velocity perpendicular to the shock front. Given the typical composition and pre-shock velocities of stellar winds, the plasma in the post-shock region is heated to temperatures of $\sim 10$\,MK, i.e.\ significantly higher than what is expected for wind-embedded shocks of single stars. What happens to this plasma in the post-shock region depends on the efficiency of radiative cooling quantified via the cooling parameter \citep{SBP}:
\begin{equation}
  \chi = \frac{t_{\rm cool}}{t_{\rm esc}}
  \label{chicooling}
\end{equation}
with $t_{\rm cool}$ and $t_{\rm esc}$ respectively the typical timescales for radiative cooling and the escape time from the shock region. If $\chi << 1$, radiative cooling is highly efficient, the post-shock material gives away a lot of energy, and its temperature quickly drops. This situation applies to close, short-period, binaries where the higher density in the interaction region renders radiative cooling more efficient. As a result, the wind interaction zone is dominated by relatively cool gas, with only little high-temperature material left. The ensuing X-ray emission is expected to scale linearly with the incoming kinetic energy flux \citep{Kee}. Conversely, if $\chi >> 1$, the shocked plasma cools adiabatically, implying that it remains much hotter. This is the case in relatively wide, long-period, CWBs with rather low pre-shock wind densities and thus lower post-shock plasma densities. Under these circumstances, the X-ray luminosity is expected to scale with $\dot{M}^2/d$ where $d$ is the orbital separation between the stars \citep{SBP,Pit18}. Whereas the post-shock regions are very thin when radiative cooling dominates, they are significantly wider in the adiabatic regime, where the shocks are typically located at an angle of $\sim 20^{\circ}$ from the contact discontinuity \citep{Pit18}.

Much insight into the theoretical properties of CWBs has been gained through numerical hydrodynamic simulations. Over the years, models of increasing sophistication were designed to account for a number of physical effects that impact the plasma properties. Indeed, several processes can lower the Mach number of the wind ahead of the shock and thus the temperature of the shocked plasma. These include inhibition of the wind acceleration and radiative braking by the companion's radiation field \citep{Ste94,Gay97}, ionization of the inflowing wind by the radiation from the post-shock plasma which lowers the efficiency of radiative acceleration of the gas ahead of the shock \citep{Par13}, thermal conduction due to electrons \citep{Mya98}, and the pressure that relativistic particles exert on the pre-shock flow in binary systems where electrons are accelerated to relativistic velocities \citep{Pit06}. The temperature of the post-shock gas can also be reduced through inverse Compton scattering of stellar photons by the electrons in the post-shock region \citep{Mya93}. Further improvements of the hydrodynamical simulations include the development of genuine 3-D adaptive-mesh models accounting for radiative driving, gravity, radiative cooling as well as orbital motion and the ensuing Coriolis deflection \citep{Par08,Pit09,Par11,Par14}.

As becomes apparent from Eq.\,\ref{kinetic}, because of their very different masses, ions should undergo a significantly stronger increase in temperature than electrons upon crossing the shocks. In a plasma where Coulomb interactions dominate, the electron and ion temperature equalize within a short time. Yet, in lower density environments this process can be slow, and evidence for non-equilibrium situations has been found in wide WR + O systems \citep[e.g.][]{Zhe00,Pol05,Zhe21}, whereas in shorter period ($< 1$\,month) WR binaries, electron and ion temperatures were found to be equal \citep{Zhe12}.\\

As discussed above, the $L_{\rm X}/L_{\rm bol}$ ratios of many O-star binaries are not significantly different from those of single O-stars. Moreover, O + O binaries do not necessarily display a hard X-ray emission \citep[e.g.][and references therein]{RauNaz16}. The same conclusion applies to WR + O systems: a survey of 20 WR + O binaries, with periods ranging from 1.75\,days to ten years, revealed clear evidence of colliding wind X-ray emission in only six systems and hints of such an emission in three other systems \citep{NGM21}. Most remarkably, five systems displayed $\log{L_{\rm X}/L_{\rm bol}} \leq -8$, indicating that the presence of a wind-wind collision is unlikely. This raises the question why some CWBs are X-ray bright whilst others are not. For systems where radiative cooling is efficient, part of the answer could come from the thin-shell instability  \citep{SBP,Kee,Ste18}. These instabilities distort the wind interaction zone into extended shear layers with oblique shocks, implying a reduction of the post-shock temperature and a loss of energy of the hot gas due to the mechanical work that it performs on the interleaved filaments of cold gas. This situation leads to a considerable reduction of the X-ray emission \citep{Kee,Ste18} compared to expectations for a steady radiative plasma.

The effect of thin-shell instabilities was directly observed in the case of the inner binary (LBV/WN + WN4, $P_{\rm orb} = 19.3$\,d, $e \simeq 0.3$) of the triple system HD5980. Based on observationally determined wind parameters, the radiative cooling and escape times were evaluated, and Eq.\,\ref{chicooling} indicates that the wind interaction of this system is in the radiative regime all around the orbital cycle \citep{Naz18b}. The observed X-ray flux scales linearly with orbital separation $d$, probably due to the fact that the winds reach higher pre-shock velocities at larger $d$ \citep{Naz18b}. The primary star of HD5980 underwent two LBV eruptions in 1993-1994. To study the evolution of the wind interaction as the primary wind is progressively returning to its pre-eruption state, the X-ray lightcurve was monitored twice: at first between 2000 and 2005 \citep{Naz07} and a second time in 2016-2017 \citep{Naz18b}. Whilst the linear scaling of the X-ray flux with $d$ was found at both epochs, the system displayed a significantly brighter and harder X-ray emission during the second epoch. This change of the overall properties likely reflects the impact of thin-shell instabilities \citep{Kee}. Indeed, though radiative cooling was dominant at both epochs, its importance had nevertheless significantly decreased in 2016-2017 compared to the earlier epoch, thereby reducing the impact of thin shell instabilities on the X-ray emission \citep{Naz18b}.\\ 

Phase-locked variability of the observed X-ray flux of colliding wind systems is expected, either as a result of the changing line of sight as the stars revolve around each other, or as a consequence of changes of the intrinsic emission due to a changing orbital separation in eccentric binaries, or as a combination of both effects \citep{Pit10}. Monitoring of the X-ray emission of colliding wind binaries as a function of orbital phase revealed a variety of such phase-locked effects \citep{RauNaz16}.

For instance, in the short-period ($P_{\rm orb} = 4.21$\,days) WN5 + O6 system V444~Cyg, eclipses of the X-ray emission region by the stellar bodies were observed, and their asymmetry was attributed to the impact of the Coriolis deflection on the wind interaction \citep{Lom15}. In a number of wider systems, the observed X-ray emission was found to be strongly attenuated at orbital phases when the star with the more powerful wind passes in front \citep{Gos09,Pan14,Gos16,Zhe21}, or alternatively to increase at phases when the star with the weaker wind passes in front \citep{Wil95}. How exactly the phenomenon manifests itself (wind eclipse or brief increase of the emission) depends on the opening angle of the shock cone (hence on the wind momentum ratio), the orbital eccentricity and the orbital orientation. Yet, in both cases, it stems from the variation of the optical depth for photoelectric absorption along the line of sight towards the wind interaction region. Such effects are most prominently seen in systems with a strong contrast between the strength of both winds such as in WR + O binaries (e.g.\ $\gamma^2$~Vel, WR~21a, WR~22, WR~25, WR~140).

Due to the phase-dependent orbital separation, wide eccentric binaries offer an ideal testbed for the theory of adiabatic wind interactions, and more specifically for the predicted dependence of the X-ray flux on $1/d$. Evidence for such variations was indeed found in several O + O and WR + O binaries \citep{Naz12b,Pan14,Gos16,NGM21},
  but strong deviations from this simple expectation were observed in other systems among which the long-period (8 -- 9\,years) binaries 9~Sgr \citep{Rau16b} and WR~140 \citep{Zhe21}. Some departures from the $1/d$ scaling could be due to radiative inhibition or braking, though the efficiency of this mechanism is reduced by the wide separation even around periastron passage. For 9~Sgr and WR~140, part of the explanation may reside in the fact that both systems host relativistic electrons in their wind interaction zone as unveiled by their synchrotron radio emission. This could lead to shock modification due to particle acceleration \citep{Pit06}. More surprisingly, whilst 9~Sgr and WR~140 still display an increase of their emission near periastron, such trends are totally absent in $\gamma^2$~Vel \citep{Rau00} and WR~22 \citep{Gos09} which have orbital periods near 80\,days. The reason for the lack of an increase of the X-ray emission at periastron in these systems is a puzzle, though it might indicate that the WR wind directly crashes onto the O-star's photosphere at all orbital phases \citep{Par11b}. 
\begin{figure}
  \begin{center}
    \resizebox{8cm}{!}{\includegraphics{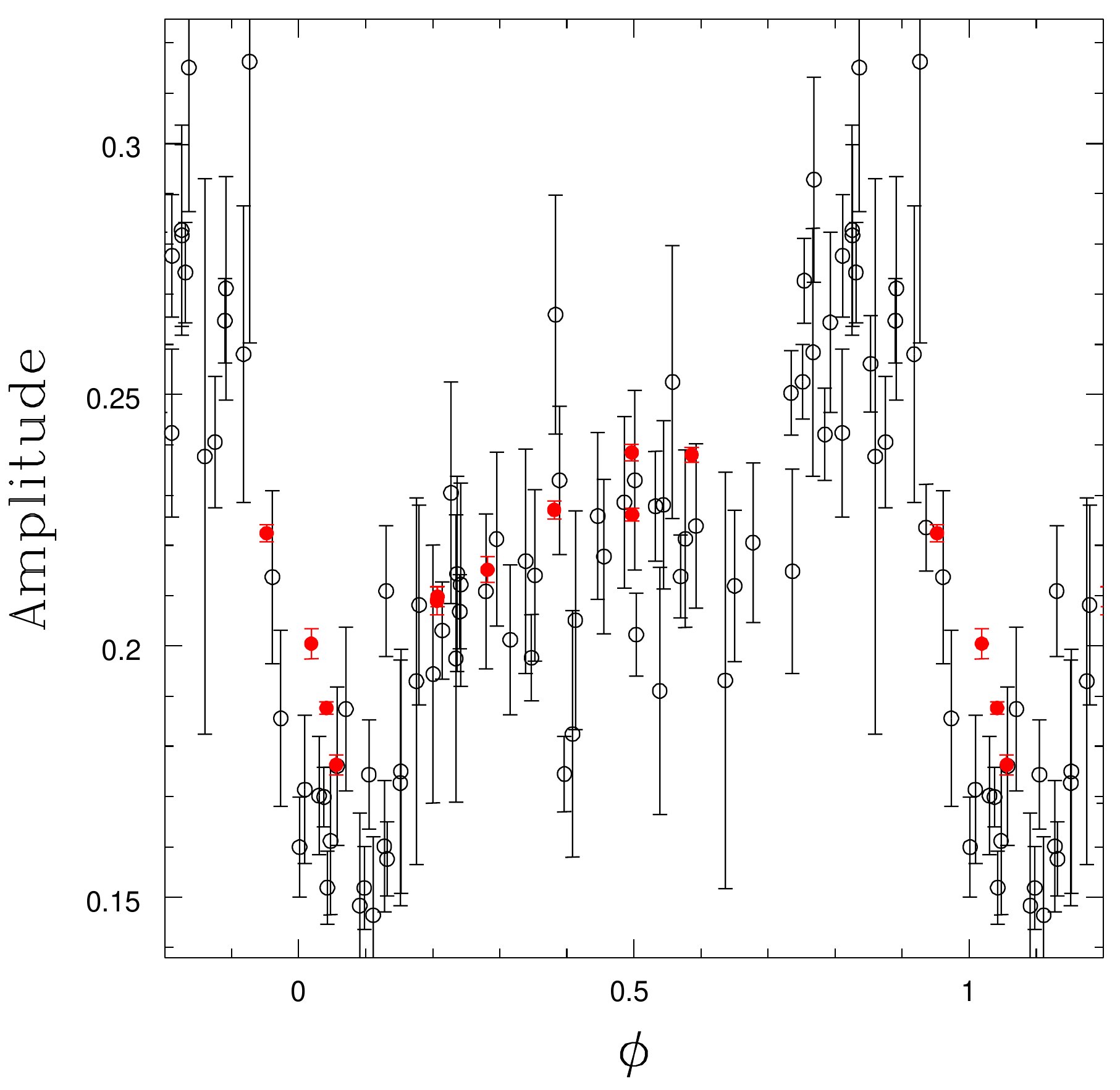}}
  \end{center}
  \caption{Orbital variations of the X-ray emission of Cyg\,OB2~\#8a \citep[O6\,I + O5.5\,III, $P_{\rm orb} = 21.9$\,d, $e = 0.18$,][]{Mos20} as observed in the 2 - 10\,keV band with {\it Swift} (open symbols) and {\it XMM-Newton} (filled red symbols). Phase 0.0 corresponds to periastron passage. The vertical axis indicates the {\it Swift} XRT count rate. The {\it XMM-Newton} EPIC-pn count rates have been scaled by a factor 0.5 to match those of {\it Swift} at the same phases. \label{Cyg8a}}
\end{figure}

In some eccentric binaries with orbital periods of the order a week to one month, the nature of the wind interaction zone (adiabatic versus radiatively cooling) changes between apastron and periastron \citep[][see the case of Cyg\,OB2~\#8a in Fig.\,\ref{Cyg8a}]{Pit10,Mos20}. In such cases, the X-ray emission at a specific orbital phase not only depends on the instantaneous separation, but also on the history of the plasma at previous phases. When the X-ray flux is plotted as a function of orbital separation, these systems display hysteresis-like loops \citep[e.g.][]{Pit10,Caz14,RauNaz16,Mos20}. Moreover, in some other systems, evidence has been found for a collapse of the wind collision region onto the surface of the star with the weaker wind around periastron, leading to a sudden and significant drop of the X-ray emission followed by a progressive recovery as the orbital separation increases again \citep{Gos16,Naz17,Pol18}.  

Non-thermal synchrotron radio emission observed in a subset of the CWBs unveils the presence of a population of relativistic electrons in the wind interaction region of these systems \citep{Ben10,Blo13}. These particles are accelerated through diffusive shock acceleration in the wind interaction region \citep{Pit21}. This opens up the possibility that inverse Compton scattering of stellar UV photons by these relativistic electrons could result in a non-thermal X-ray or $\gamma$-ray emission \citep[e.g.][]{Che91,Pit06,Rei06,Pit20,Pit21}. Still, so far all attempts to detect such a non-thermal X-ray emission in the 0.5 -- 10\,keV X-ray band failed, indicating that any non-thermal X-ray emission must be significantly weaker than the thermal emission from CWBs. However, the $\eta$~Car CWB system was detected at energies above 20\,keV with {\it INTEGRAL} and {\it NuSTAR} \citep{Ley10,Ham18} and up into the GeV domain with {\it Agile} and {\it Fermi} \citep{Tav09,Abdo,Rei15}. This suggests that other CWB systems could be detected as hard X-ray or even $\gamma$-ray sources provided they would be observed with a sufficient sensitivity in an energy domain where the thermal emission is negligible. Yet, {\it INTEGRAL} and {\it NuSTAR} observations of the Cyg~OB2 association, which hosts several non-thermal radio emitter O + O binaries, failed to detect non-thermal X-ray emission from those systems \citep{Mos20}. It seems thus that for the majority of the CWBs, any putative non-thermal emission component has an intensity below the sensitivity limit of current instrumentation.\\       

\begin{figure}
    \begin{center}
    \resizebox{11cm}{!}{\includegraphics{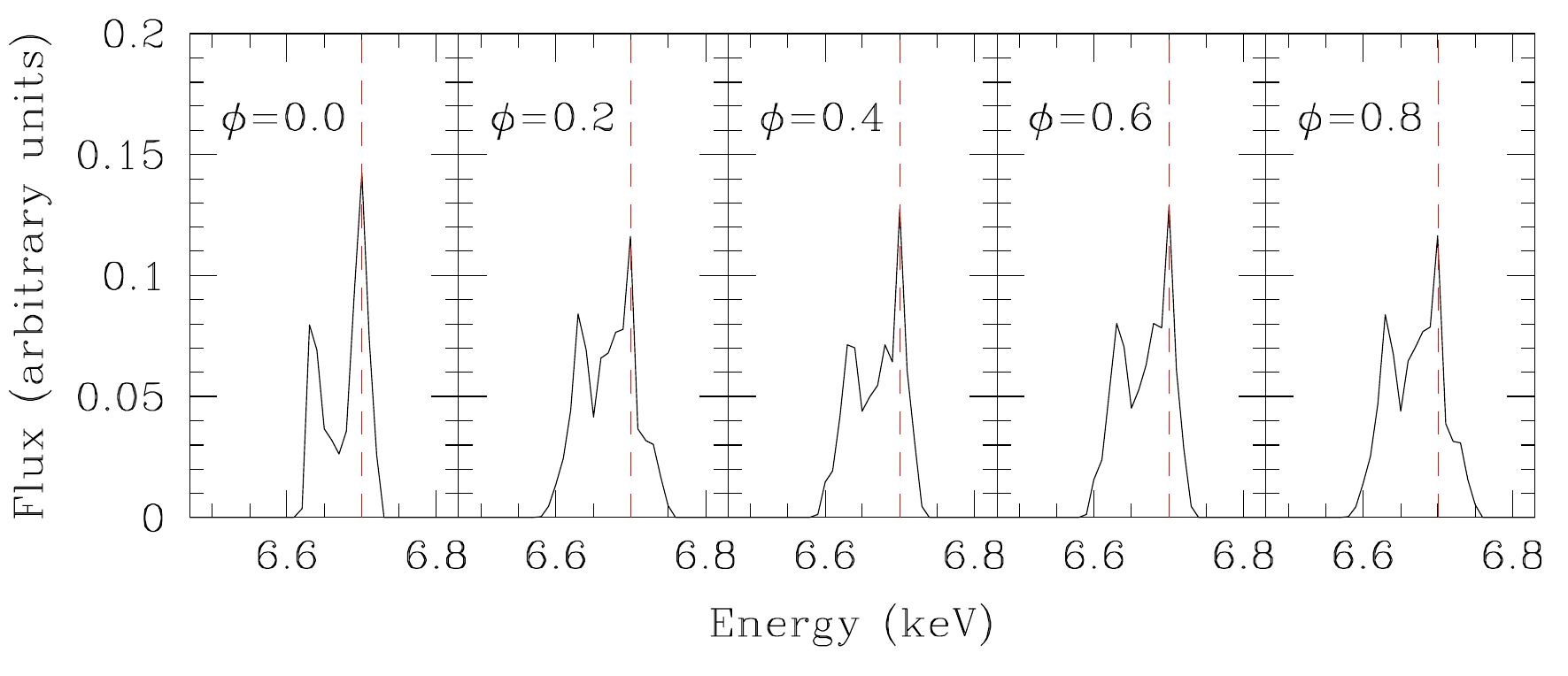}}
    \end{center}
  \caption{Synthetic profiles of the Fe\,{\sc xxv} triplet at 6.7\,keV for a colliding wind binary consisting of an O5\,I primary and an O3\,III secondary with an orbital separation of 564\,R$_{\odot}$, seen under an inclination of $i=72^{\circ}$ at 5 orbital phases. Profiles were computed with the Lifeline code \citep{Mos21}. The wind interaction zone is in the adiabatic regime. The dashed red vertical line indicates a photon energy of 6.7\,keV. Phase 0.0 is defined as the conjunction with the O5\,I star in front.\label{Lifeline}}
\end{figure}

Compared to single O-type stars, there is much less literature on high-resolution X-ray spectroscopy of CWBs. Theoretical predictions of such line profiles and their orbital variability have been made either based on 2-D hydrodynamic simulations of adiabatic wind collisions \citep{Hen03} or using instead semi-analytical formalisms \citep{Can96,Ant04} for adiabatic or steady radiative wind interactions \citep{RMN16,Mos21}. Ref.\,\citep{Hen03} focused on the Ly$\alpha$ lines of O\,{\sc viii}, Ne\,{\sc x}, Mg\,{\sc xii}, Si\,{\sc xiv} and S\,{\sc xvi}. Yet, in most CWBs, these lines are, at least partially, emitted also by the individual winds. The essentially unknown contributions from the individual winds could thus blur the picture, except for systems such as $\eta$~Car \citep{Hen08} or WR~140 \citep{Pol05,Zhe21} where the colliding wind emission overwhelms the intrinsic emission from the individual winds. Subsequent studies \citep{RMN16,Mos21} instead concentrated on synthetic line profiles of the He-like triplet of Fe\,{\sc xxv} near 6.7\,keV. The emission of the latter lines requires plasma temperatures that are usually not reached in the winds of single non-magnetic massive stars. Hence, they should offer a direct diagnostic of the colliding wind interaction. Whilst the Fe\,{\sc xxv} 6.7\,keV line profiles cannot be studied in detail with the grating spectrographs onboard of {\it Chandra} and {\it XMM-Newton}, future bolometric facilities such as Resolve on {\it XRISM} \citep{Tas20} and X-IFU on {\it Athena} \citep[][see also chapter by Guainazzi, Nandra \& Barret]{Bar20} will be ideally suited for such investigations (see Fig.\,\ref{Lifeline}).

Meanwhile, existing high-resolution spectra of CWBs revealed some discrepancies with the theoretical expectations. For instance, in the case of $\gamma^2$~Vel, the lines observed shortly after the passage of the O-star in front of the WC8 star showed no blueshift, whilst a $-300$\,km\,s$^{-1}$ blueshift was expected \citep{Hen05}. This could hint at a wide opening angle ($\sim 85^{\circ}$) of the shock cone \citep{Hen05}, though such a wide angle is at odds with the brevity of the increase of X-ray flux when the cone sweeps across our line of sight \citep{Wil95}. In $\eta$~Car, high-resolution HETG spectra collected over one half of the orbital cycle revealed clear changes of the Si\,{\sc xiv} and S\,{\sc xvi} line centroids. These variations could only be reconciled with the constraints derived from the broadband X-ray lightcurve assuming a specific dependence of the emissivity distribution along the shock front \citep{Hen08}. Similarly, HETG spectra of WR~140 revealed line profiles that were generally not well matched by synthetic profiles \citep{Zhe21}. 

\section{$\gamma$~Cas stars}
For over 150 years, $\gamma$~Cassiopeiae (B0.5\,IVpe) has been known as the prototype of the so-called Be stars, i.e.\ main-sequence or giant B-type stars displaying bright, often double-peaked, Balmer hydrogen emission lines in their optical spectrum. These features are found in about 20\% of the B-type stars, and are interpreted as the signature of a viscous near-equatorial circumstellar decretion disk in keplerian rotation and consisting of material ejected by the star \citep[][and references therein]{Riv13}. Although the mechanism by which the material is ejected into the disk is not entirely clear yet, it is most likely related to the fast, near critical, rotation of these stars, possibly coupled to pulsations. What makes $\gamma$~Cas special, is that is has now also become the prototype of a class of mysterious X-ray sources, the so-called $\gamma$~Cas stars or $\gamma$~Cas analogs \citep{Smi16,NazMot,Naz20a}. Objects of this class display X-ray properties that distinguish them from those of other Be stars:
\begin{enumerate}
\item Their X-ray luminosities ($4\,10^{31}$ -- $2\,10^{33}$\,erg\,s$^{-1}$) are between a factor 10 and 100 higher than those of ordinary Be stars \citep{NazMot}, but still about 30 times lower than the low state of the faintest confirmed Be High-Mass X-ray binaries (Be-HMXBs) \citep{Smi98a,Reig}. The $L_{\rm X}/L_{\rm bol}$ ratios of $\gamma$~Cas stars reach values up to $10^{-5}$.
\item Their X-ray spectra are thermal as revealed by the Fe\,{\sc xxv} and Fe\,{\sc xxvi} emission lines at 6.7 and 6.97\,keV \citep{Mur86}, but with much higher plasma temperatures ($kT > 5$\,keV, often around 10 --  15\,keV) than usually seen in massive stars \citep{Smi04,Lop07,LSM10}. {\it INTEGRAL}, {\it Suzaku} and {\it NuSTAR} observations of $\gamma$~Cas and its brightest analog HD110432 (B0.5\,IVpe) revealed that the emission of this very hot plasma extends up to energies of 50 -- 100\,keV \citep{Shr15,Tsu18}. Quite remarkably, there is no evidence for an additional non-thermal component \citep{Shr15}. Beside this dominant hot plasma, warm plasma components (with $kT$ in the range 0.1 -- 3\,keV) contribute in some cases to the emission in the 0.5 --  10\,keV energy band \citep[][see Fig.\,\ref{gCasRGS}]{Smi04,Lop07,LSM10,Tor12}. The {\it f i r} triplets of helium-like ions display $f/i$ ratios indicating that they either arise in a high-density environment or close to a strong UV source \citep{Smi04,Tor12}. The X-ray lines are broad with widths of $\sim 400$ -- $500$\,km\,s$^{-1}$ for $\gamma$~Cas \citep{Smi04,LSM10} and $\sim 1200$\,km\,s$^{-1}$ for HD110432 \citep{Tor12}.
\begin{figure}
    \begin{center}
    \resizebox{10cm}{!}{\includegraphics{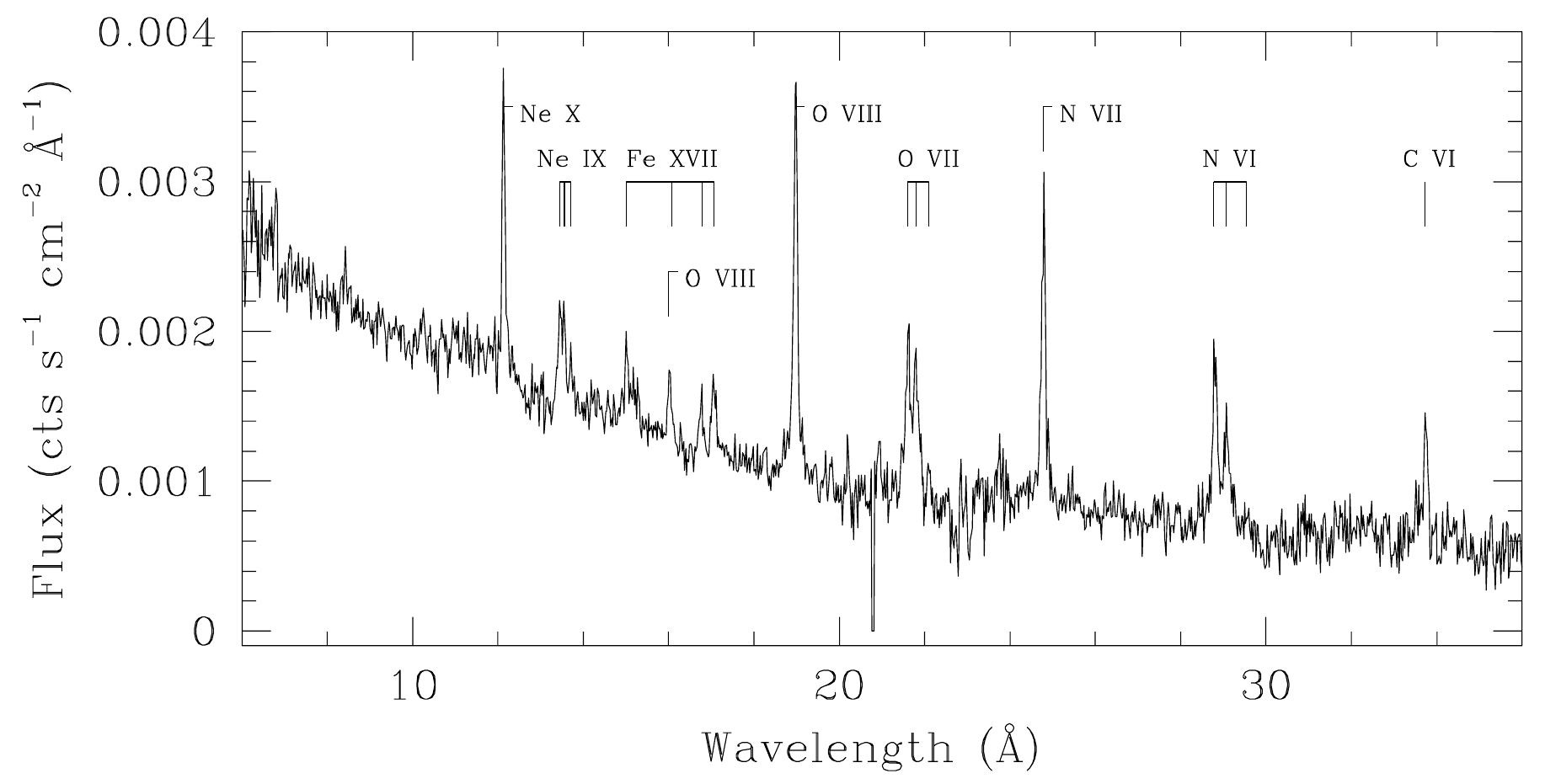}}
    \end{center}
  \caption{Combined {\it XMM-Newton} RGS\,1 and 2 spectrum of $\gamma$~Cas as observed in February 2004. The most prominent emission lines are labelled. Note the increase of the continuum level towards shorter wavelengths due to the presence of the hot ($kT \simeq 12$\,keV) plasma component. \label{gCasRGS}}
\end{figure}
\item Their X-ray emission is variable over a wide range of timescales, but lacking well-defined periodicities \citep{Par93}. The most intriguing feature of this variability is the presence of rapid shots \citep{Smi98a,Smi98b,Smi99}, occurring on timescales as short as the 4\,s time resolution of the {\it RXTE}-PCA measurements \citep{Smi98a}. The frequency of shots with a specific integrated energy decreases exponentially with energy \citep{Rob00}. Overall, the X-ray emission consists of these rapid shots superimposed on a more slowly varying basal emission. Quite remarkably, the basal and shot components have similar hardness and the hardness does not vary during the shots.
\item Those $\gamma$~Cas stars that are sufficiently X-ray bright to allow collecting high-quality spectra display a weak, but definite, fluorescent Fe K$\alpha$ line near 6.4\,keV arising from low ionization stages of iron. The EW of these fluorescent lines ranges between $\sim 30$ and 170\,eV \citep{Gim15}. 
\end{enumerate}

Over the last decade, the number of known $\gamma$~Cas stars has increased significantly: currently, there are 25 confirmed members and two additional candidates \citep{Naz20a}. All have spectral types between O9e and B3e. More members are likely to be identified in the near future with the {\it eROSITA} All-Sky Survey\citep{Predehl}. Most interestingly, Be stars seem to display a continuum in their X-ray properties (luminosity and hardness) ranging from objects displaying a normal, soft emission to the most extreme hard and bright $\gamma$~Cas spectra \citep{Naz20a}. The $\gamma$~Cas phenomenon is thus not restricted to one or two stars, and its interpretation could have a significant impact on our understanding of Be stars. The scenarios proposed to explain the unusual X-ray properties can be divided into two big families depending on whether they require binarity or not. The binary scenarios in turn can be split into subcategories depending on whether the companion of the Be star is an accreting compact object, either a neutron star or a white dwarf, or a hot subdwarf. 

From an evolutionary point of view, a promising scenario to explain the near critical rotation of Be stars is by means of a past mass and angular momentum transfer episode in a binary system. If the binary system survives the subsequent evolution of the mass donor, one can expect the mass gainer, i.e.\ the current Be star, to be bound to either a compact companion (black hole, neutron star or white dwarf) or to a hot stripped helium subdwarf \cite[sdO,][and references therein]{ShaoLi2}. From the observational point of view, $\gamma$~Cas was indeed shown to be a binary with an orbital period of 203.5\,d and a nearly circular orbit \citep[$e \leq 0.03$,][and references therein]{Nem12,Smi12}. Whilst the nature of the unseen secondary star remains unknown, its likely mass (0.8 -- 1.0\,M$_{\odot}$) is compatible with either a late-type main-sequence star, an sdO star or a white dwarf. Additional $\gamma$~Cas stars were found to be binaries, suggesting that about 70\% of them could be binaries \citep{Naz22}. Whilst further studies are clearly needed, existing results seem to favor companion masses compatible with either a white dwarf or an sdO star, but not with a neutron star.    

\subsection{Accreting compact companion scenarios}
The first scenario that was proposed was that $\gamma$~Cas could be an HMXB where the X-rays result from accretion of material onto a neutron star companion \citep{Whi82}. Yet, despite intensive searches, no X-ray pulsations - typical for accretion onto a spinning neutron star - were found \citep{Whi82,Par93}. Beside the lack of pulsations, the X-ray emission of $\gamma$~Cas stars differs from that of Be-HMXBs in terms of its luminosity, variability (absence of bursts) and its thermal nature. These differences led to the abandon of the accreting neutron star scenario, until \citet{Pos17} argued that $\gamma$~Cas might include a fast spinning magnetized neutron star. For such objects, spherical accretion would be impeded by the propeller mechanism, leading to the formation of a hot shell of material around the neutron star magnetosphere that would emit thermal X-rays and not produce pulses \citep{Pos17}. Whilst the propeller scenario addresses some of the earlier objections against an accreting neutron star companion, it still faces a number of problems \citep{Smi17}. For instance, it does not account for the observed correlations between optical/UV and X-ray variations \citep{Smi98b,Mot15}. Furthermore, the Be star mass-loss rate required to feed a neutron star in the propeller regime through wind accretion would exceed the expected value by several orders of magnitude \citep{Smi17}. Finally, from the evolutionary point of view, a propeller phase would be too short to account for the growing number of $\gamma$~Cas stars and their proportion with respect to the number of known Be-HMXBs in a distance-limited sample \citep{Smi17}.

Because of the difficulties of the accreting neutron star scenario, its was suggested that the X-ray emission rather stems from accretion onto a white dwarf \citep{Mur86,Haberl,Tsu18}. Accreting white dwarfs are best known in association with cataclysmic variables and symbiotic stars, and their X-ray emissions display a wide range of properties \citep{Muk17}. Evolutionary models also predict the existence of white dwarfs gravitationally bound to Be stars. Yet, the majority of these systems should be associated to late-type Be stars, whilst the known $\gamma$~Cas stars are early-type Be stars \citep{ShaoLi}. Observationally, {\it INTEGRAL} data of $\gamma$~Cas, collected over nine years, reveal no variability of the flux at energies above 20\,keV, unlike accreting white dwarf systems that do exhibit variability at these energies \citep{Shr15}. Analyses of the X-ray spectra of $\gamma$~Cas and HD110432 with models for magnetic and non-magnetic accreting white dwarfs yield constraints on the white dwarf masses ($\sim$ 0.7 -- 0.8\,M$_{\odot}$) and accretion rates \cite[$\dot{M}_{\rm X} \sim$ 1 -- 2\,$10^{-10}$\,M$_{\odot}$\,yr$^{-1}$,][]{Tsu18}. These accretion rates would either imply a mass-loss rate from the Be star several orders of magnitude higher than the observationally inferred wind mass-loss rate, or imply that the white dwarf must orbit the Be star in the plane of its decretion disk. Another issue is once again, that this scenario cannot explain the observed correlations between optical/UV and X-ray variations of $\gamma$~Cas \citep{Smi98b,Mot15}.

\subsection{Hot subdwarf companion scenario}
As pointed out above, binary evolution could lead to the formation of systems consisting of a Be star orbited by a hot stripped helium star subdwarf. Since such hot subdwarfs have stellar winds, it was proposed that the hard X-ray emission of $\gamma$~Cas stars arises from the collision of the wind of an sdO companion with the Be disk and/or wind \citep{Lan20}. A key point here is whether or not the power released in such a wind -- disk interaction would be sufficient to account for the observed X-ray luminosity of $\gamma$~Cas stars. Wind parameters have been determined observationally for a sample of hot, presumably single, extreme helium stars \citep{JH10}, as well as for HD45166 which contains a `quasi-WR' star orbiting a B7\,V companion \citep{Groh}. Terminal wind velocities of the single sdO stars range between 400 and 2000\,km\,s$^{-1}$ \citep{JH10}. Likewise, the terminal wind velocity of HD45166 (T$_{\rm eff} = 50$\,kK) is only 350\,km\,s$^{-1}$ \citep{Groh}. These values are much smaller than the 3000 to 5000\,km\,s$^{-1}$ adopted by \citep{Lan20}. Mass-loss rates determined from the spectral analysis of single sdO stars range between $\log{\dot{M}} = -7.3$ and $-9.8$ (M$_{\odot}$\,yr$^{-1}$), more massive helium stars having larger $\dot{M}$ \citep{JH10}.  Yet, only the lowest sdO masses (and thus the lowest $\dot{M}$ values) are compatible with the dynamical mass determinations discussed above. The most realistic estimates of the He star wind luminosities are thus significantly lower than assumed by Ref.\,\citep{Lan20}. Moreover, only a fraction of the sdO wind mechanical luminosity (corresponding to the part of the sdO wind that is shocked by the Be disk or wind) would actually be converted into heat. Based on observed sdO wind properties, the luminosity that could possibly be released by this mechanism is thus several orders of magnitude lower than what would be needed to explain the $\gamma$~Cas phenomenon. The maximum postshock plasma temperature of such an interaction can be estimated from Eq.\,\ref{kinetic}. For a given wind velocity, this yields a maximum postshock plasma temperature a factor $\sim 4$ lower than quoted by Ref.\,\citep{Lan20}. Further accounting for the overestimate of $v_{\infty}$ discussed above, it seems unlikely that the collision of the sdO wind with the Be disk or wind results in a plasma as hot as observed in $\gamma$~Cas stars. 

\subsection{Magnetic star/disk interaction}
Multi-wavelength campaigns showed that the variability of the X-ray emission of $\gamma$~Cas is strongly correlated to its UV and optical variability on timescales of hours and days \citep[][and references therein]{Smi99}. These correlations actually hold over significantly longer timescales. Indeed, X-ray data of $\gamma$~Cas, obtained with the {\it RXTE} all-sky monitor (1.5 -- 12\,keV) between 1996 and 2010 and with the {\it MAXI} all-sky imager (2.0 -- 20\,keV) between 2009 and 2013, are well correlated to the variations of the optical emission with a $3\,\sigma$ upper limit of one month on the delay between the variations in both wavelength ranges \citep{Mot15}. Since the migration time of material from the inner parts of the disk (where the $V$-band flux arises) to the orbit of the companion is estimated to several years, these correlations cannot easily be explained in the context of the companion scenarios. This led to the formulation of an alternative scenario where the X-ray emission is the result of magnetically-generated structures near the stellar surface \citep{Smi98a,Rob00,Mot15,Smi19}. The magnetic star/disk interaction scenario assumes that small-scale magnetic fields emerging from the Be star entangle with a toroidal magnetic field generated by a dynamo mechanism in the inner parts of the Be disk. To date no large-scale magnetic fields have been detected for Be stars \citep{Wad16}. This is not surprising since such large-scale magnetic fields should actually destroy a keplerian circumstellar disk \citep{udD18}. The stellar magnetic fields required for the star-disk interaction scenario would instead be small-scale localized fields generated in the thin subsurface convective layers due to the iron opacity peak \citep{Can09}. Owing to the different rotation rates of the two systems of magnetic field lines, the lines stretch and eventually sever, leading to magnetic reconnection events that accelerate electron beams onto the Be star's surface where they thermalize and expand explosively \citep{Smi19}. These impacts would produce the rapid shots. Only a fraction of the energy is radiated during the shot event, the remainder of the hot plasma is expected to expand adiabatically along the magnetic loop, thereby producing a corona of hot plasma responsible for the basal emission \citep{Smi19}.

\begin{figure}
  \begin{minipage}{6.5cm}
    \begin{center}
    \resizebox{6.5cm}{!}{\includegraphics{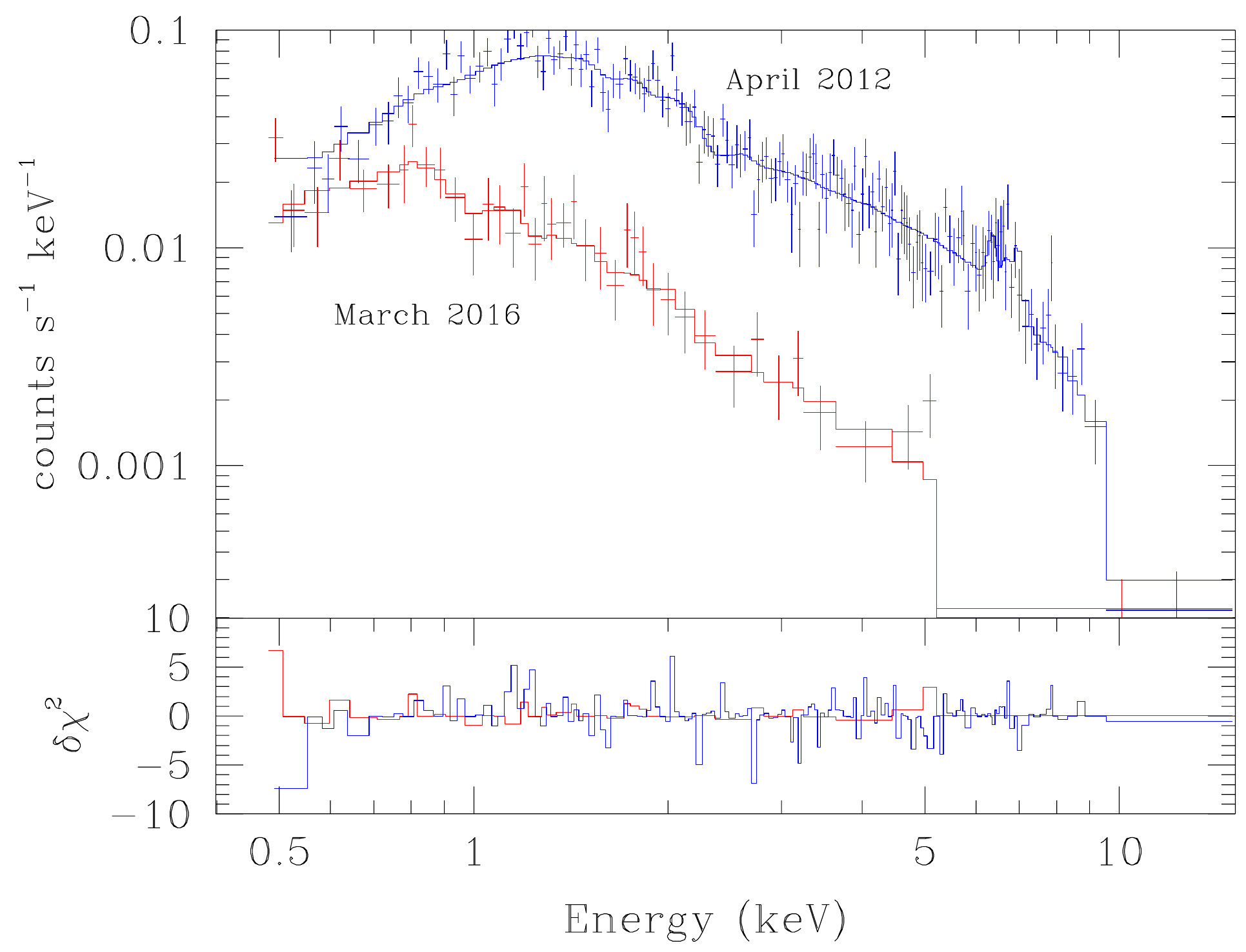}}
    \end{center}
  \end{minipage}
  \begin{minipage}{5cm}
    \begin{center}
    \resizebox{5cm}{!}{\includegraphics{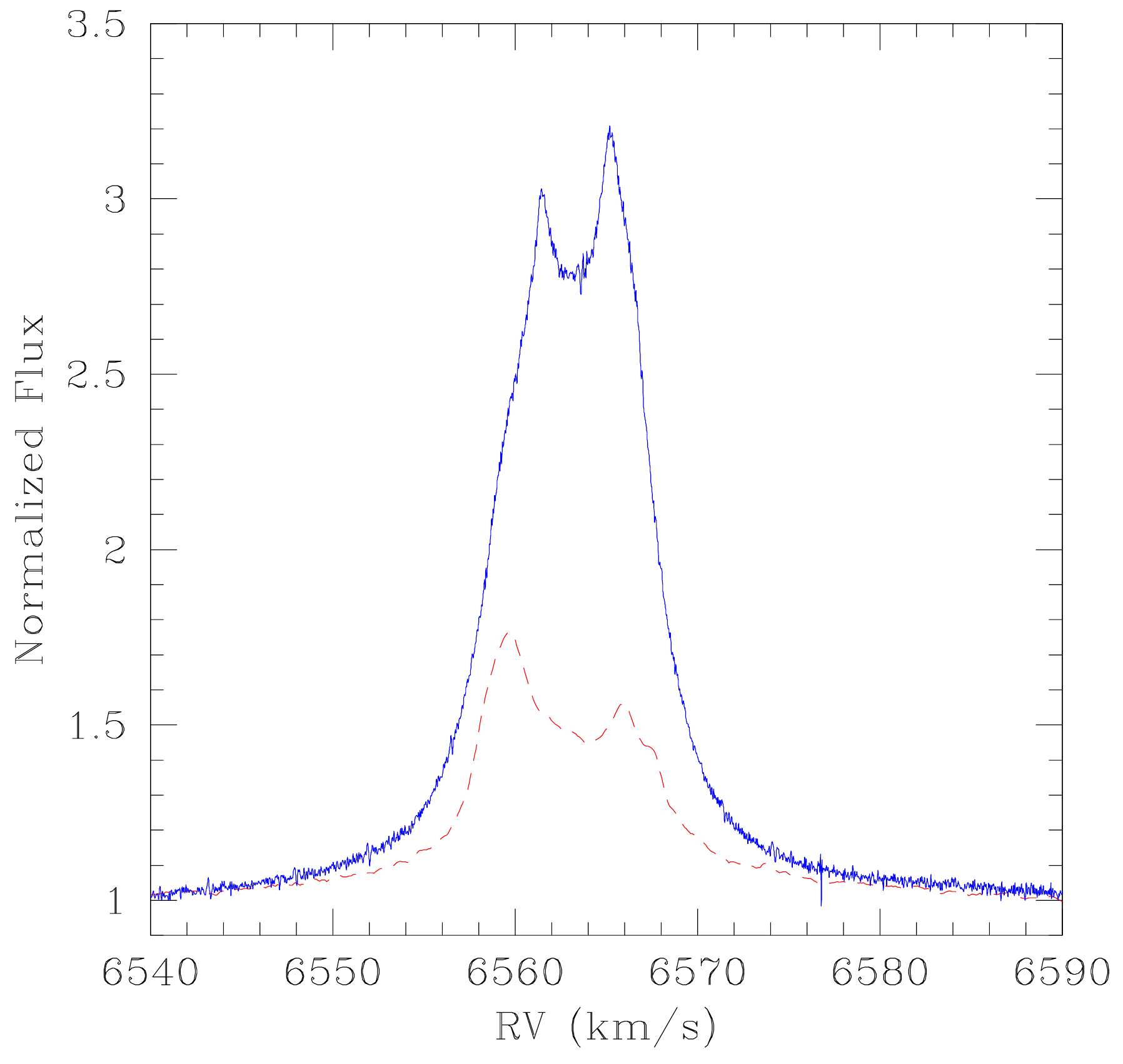}}
    \end{center}
  \end{minipage}  
  \caption{Left: variations of the {\it XMM-Newton} EPIC-pn spectrum of HD45314 between April 2012 (blue) and March 2016 (red). The top panel illustrates the spectrum,, whilst the lower panel shows the contributions of the various energy bins to the $\chi^2$ of the best-fit model \citep{Rau18}. Right:  variations of the H$\alpha$ line between April 2012 (blue) and March 2016 (red).\label{HD45314}}
\end{figure}

Be disks are known to be strongly variable, and can sometimes even dissipate. Whilst all the scenarios introduced above are expected to produce an X-ray emission that somehow scales with the density of the Be disk, the magnetic interaction scenario is clearly the one where the X-ray emission is expected to react fastest to a disk outburst or dissipation. With this idea in mind, several $\gamma$~Cas stars have been monitored in the optical domain and X-ray observations were triggered when the disk conditions were changing.
In 2014, the O9e star HD45314 entered a phase of spectacular variations with the optical emission lines undergoing important changes of their morphology and strength. This included transitions between single- and multiple-peaked emission lines as well as shell events, and a phase of (partial) disk dissipation \citep{Rau18}. The star was observed three times in X-rays. A first {\it XMM-Newton} observation, taken in April 2012, prior to the onset of the spectacular variations when the H$\alpha$ emission was strong (EW(H$\alpha$) = -22.6\,\AA), revealed the $\gamma$~Cas nature of this star. A {\it Suzaku} observations was obtained during the shell phase in October 2014, and a second {\it XMM-Newton} observation during partial disk dissipation (EW(H$\alpha$) = -8.5\,\AA) \citep{Rau18}. Whilst HD45314 preserved its hard and bright X-ray emission during the shell phase, the October 2014 X-ray spectrum was significantly softer and weaker (see Fig.\,\ref{HD45314}), suggesting a direct association between the level of X-ray emission and the disk density.
A contrasting picture was obtained for $\pi$~Aqr \citep[B1\,Ve,][]{NRS19}. A coordinated X-ray ($\sim 1$\,ks snapshots with {\it Swift}) and optical monitoring was conducted over three 84\,d orbital cycles in 2018. At that time, the star displayed a very strong H$\alpha$ emission (EW(H$\alpha$) between $-22$ and $-25$\,\AA). The X-ray emission was found to be variable, but without any obvious correlation with orbital phase or H$\alpha$ emission strength. The average X-ray flux and the relative amplitude of the recorded flux variations were similar to those seen in a single 55\,ks {\it XMM-Newton} observation obtained in 2013 when the disk had nearly disappeared (EW(H$\alpha$) = -1.7\,\AA). This suggests that the mean level of the X-ray flux remained basically unchanged between 2013 and 2018, despite the huge change in H$\alpha$ emission strength between those epochs \citep{NRS19}. Further monitorings of $\gamma$~Cas stars are needed to assess the connection between the Be-disk and the X-ray emission in more details.  

\section{Conclusions and future prospects}
Our understanding of the X-ray properties of massive stars has been tremendously improved by the fleet of past and current X-ray satellites that allowed us to study the spectra and lightcurves of a number of objects. Yet, as highlighted in this chapter, there are still a number of open issues that need to be solved. These topics will be addressed with the upcoming facilities, notably the Resolve and X-IFU cryogenic micro-calorimeters on board the {\it XRISM} and {\it Athena} missions \citep{Tas20,Bar20}. The capabilities of these next-generation facilities will allow to fully exploit the diagnostic potential of high-resolution spectroscopy around the Fe K lines and of time-resolved high-resolution spectroscopy \citep{Sci13}. This will open new territories in the studies of massive stars, allowing for instance to perform detailed comparisons of observed line morphologies with model predictions for single massive stars, colliding wind binaries and $\gamma$~Cas stars. Time-resolved spectroscopy will notably allow to perform Doppler tomography of the emission regions in colliding wind interactions, and to probe the behaviour of individual emission lines in pulsating massive stars. These studies will shed new light on the dynamics and driving mechanisms of stellar winds, and help solve the mystery of $\gamma$~Cas stars. There definitely is an exciting future ahead for the study of X-ray emissions in massive stars.

\section*{Acknowledgments} I'm grateful to Ya\"el Naz\'e for discussion. This work was supported by the Fonds National de la Recherche Scientifique under grant PDR T.0192.19 and by the Belgian Federal Science Policy Office (Belspo) through the HERMeS PRODEX contract.

\section*{Cross references}
\begin{itemize}
\item[$\bullet$] Wilkes \& Tananbaum (chapter on Chandra mission), referenced on page 2
\item[$\bullet$] Schartel \& Santos-Lle\'o (chapter on XMM-newton  mission), referenced on page 2
\item[$\bullet$] ud-Doula \& Owocki (chapter on magnetically-confined wind shocks), referenced on page 12, first paragraph of Sect. 2.3 and middle of page 12
\item[$\bullet$] Guainazzi, Nandra \& Barret (chapter on Athena mission), referenced on page 18
\end{itemize}

\end{document}